# Navigating Information and Uncertainty :

## A Fuzzy Logic Model to Approach Transparency, Democracy and Social Wellbeing

**Carlos Medel Ramírez**
ORCID: 0000000256416270
IIESES / Universidad Veracruzana

**Hilario Medel López**
ORCID 0000000200728654
IA / Universidad Veracruzana

**Jennifer Lara Mérida**
ORCID 0000000321215652
ED / Universidad de Xalapa

# Navigating Information and Uncertainty:

## A Fuzzy Logic Model to Approach Transparency, Democracy and Social Wellbeing

Carlos Medel Ramírez,
ORCID: 0000000256416270
Investigador del Instituto de Investigaciones y Estudios
Superiores Económicos y Sociales
Universidad Veracruzana
Coordinador del Observatorio ObservesIIESES UV
Autor de correspondencia: Email: cmedel@uv.mx

Hilario Medel López,
ORCID 0000000200728654
Investigador del Instituto de Antropología
Universidad Veracruzana

Jennifer Lara Mérida
ORCID 0000000321215652
Universidad de Xalapa
Escuela de Derecho

## List of Figures

September 2023

# Introduction

As we embark on a profound intellectual journey within the pages of Navigating Information and Uncertainty: A Fuzzy Logic Model to Approach Transparency, Democracy and Social Wellbeing (tDTSW) we stand on the threshold of a captivating exploration. This book delves deep into the intricate interplay among democracy, transparency, and social well-being, providing a distinctive perspective anchored in the realm of fuzzy logic.

Our expedition commences by recognizing the intricate nature of the contemporary era, characterized by an overwhelming deluge of information and an enduring state of uncertainty. In an age marked by digital transformation and global interconnectedness, the dynamics of governance, accountability, and societal advancement have grown increasingly convoluted. We navigate through an expansive sea of data, where clear distinctions often blur into ambiguity, and certainty yields to doubt.

The title of this book encapsulates the essence of our voyage. It acknowledges that in today's world, the pursuit of transparency, the practice of democracy, and the realization of social well-being necessitate a departure from conventional approaches. They demand a fresh perspective capable of navigating the vast ocean of information while grappling with the inherent uncertainty that accompanies it.

Our guiding beacon through this odyssey is the tDTSW model a robust framework founded on the principles of fuzzy logic. This model transcends the confines of conventional binary thinking, enabling us to embrace the nuanced, imprecise nature of reality. It furnishes a structured methodology for comprehending and analyzing the intricate interplay among democracy, transparency, and social well-being in a world where well-defined boundaries are often elusive.

In the ensuing chapters, we shall embark on a journey of profound discovery. We will explore the pivotal roles that transparency and democracy play as the cornerstones upon which equitable, just, and sustainable societies are erected. Taking inspiration from nations such as Finland, Singapore, and New Zealand, we shall witness the transformative influence of transparent and accountable governance. These case studies stand as beacons of optimism, illustrating that the cherished principles we hold dear can indeed yield tangible benefits and constructive change.

Nonetheless, our journey shall not shrink from the complexities and challenges that contemporary democracies confront. We shall confront the tensions between capitalism and ecological sustainability, grapple with matters of gender equality and delve into the vital role of education in nurturing informed and engaged citizens.

In the subsequent chapters, the tDTSW model shall unveil its all-encompassing perspective on the intricate dynamics at the intersection of democracy, transparency, and social well-being within the broader context of politics and society. It shall emphasize that democracy is not merely a theoretical concept but a potent catalyst for constructive transformation. Furthermore, it shall underscore the paramount importance of transparency, accountability, trust, and effective governance in enhancing the overall social wellbeing of citizens.

Step by step, we shall delve deeper into the world of fuzzy logic, a world that empowers us to navigate the uncertainties that often obscure the path to societal progress. We shall witness how the tDTSW model, replete with its fuzzy sets, membership functions, and fuzzy rules, furnishes a structured approach for comprehending the intricate relationship among democracy, transparency, and social well-being across diverse scenarios.

In the concluding chapters, we shall examine the pragmatic ramifications of our expedition. We shall unveil the varying degrees of influence that these variables exert on one another, bestowing upon policymakers the tools to make well-informed decisions, evaluate prevailing policies and pinpoint regional priorities for enhancing social well-being.

As we draw the curtains on this introduction, it becomes unmistakably clear that democracy and transparency are not mere abstract ideals but actionable principles capable of shaping the destiny of societies. The tDTSW model, anchored in the realm of fuzzy logic, offers a novel perspective on how to approach these principles in a world marked by information saturation and uncertainty.

Our journey toward more equitable, sustainable, and just societies necessitates dedication, adaptability, and collective effort. Governments, civil society, the private sector, and every individual have a collective responsibility in championing these ideals as the lodestar that illuminates the path to a more radiant and all-encompassing tomorrow. Our publication serves not merely as a set of directions; rather, it extends a beckoning call to partake in an epochal expedition of understanding and proactive engagement.

# Chapter 1
# Charting the Path to Social Wellbeing: The Quest for Transparency and Democracy

The tireless search for transparency and democracy is paramount in the collective desire to build fairer, more equitable and sustainable societies.[1] This principle is the pillar upon which a strong and resilient democracy is built, an indispensable element to guarantee the comprehensive wellbeing of populations. In the forthcoming chapter, a deep exploratory journey is undertaken regarding the multiple challenges and opportunities that emerge in the environment of democratic governance, accountability and the active and conscious participation of citizens.[2]

The exposition is particularly focused on how these crucial elements can, together, significantly contribute to the improvement of multidimensional social wellbeing. It reveals how, through transparency and efficient and honest public management, it is possible to advance towards the achievement of tangible social justice and guarantee[3], at the same time, environmental sustainability, an increasingly crucial aspect in the current global panorama.

Key dimensions of transparency are also addressed, breaking down its relevance and the tangible benefits it brings to society.[4] Its implications in citizen trust, in promoting a culture of responsibility and in building a more robust and healthy social fabric are discussed. In turn, the importance of implementing principles of effective democracy is addressed, which, beyond discourse, allows authentically inclusive and sustainable development for all sectors of society.

Furthermore, the text provides a thorough analysis of innovative public policies that, through their implementation, are achieving concrete and notable progress in terms of wellbeing, equity, and environmental sustainability. Successful examples are highlighted, and cases are presented that serve as reference and stimulus to replicate similar initiatives in different socio-political contexts and realities.

The chapter not only offers a complete and detailed panorama on these crucial issues, but also invites profound reflection on the still pending challenges in this important search. Lessons learned, obstacles faced and possible ways to overcome them are highlighted, in an inclusive dialogue involving governments, the private sector, civil society and the general citizen.

---

[1] See. Liubchenko, O. (2022). The general characteristics of external manifestation of procedural component verification of decisions by discussion as fundamental principle of decision-making by the parliament of Ukraine. International scientific journal "Internauka". Series: "Juridical Sciences". Series: "Juridical Sciences" https://doi.org/10.25313/2520-2308-2022-12

[2] See. Veidemane, A. (2022). Veidemane, A. Education for Sustainable Development in Higher Education Rankings: Challenges and Opportunities for Developing Internationally Comparable Indicators. Sustainability. 2022, 14, 5102. https://doi.org/10.3390/su14095102.

[3] See. Hromovchuk, M., & Byelov, D.M. (2021). The principle of humanism as a fundamental principle of building a modern rule of law. Uzhhorod National University Herald. Series: Law.

[4] See. Montero, A.G., & Blanc, D.L. (2019). The Role of External Audits in Enhancing Transparency and Accountability for the Sustainable Development Goals. UN Department of Economic and Social Affairs (DESA) Working Papers, 28 feb 2019, 29 pages DOI: https://doi.org/10.18356/3fe94447-en

In this sense, it reaffirms the premise that democratic strengthening and transparency are not just goals to be achieved, but ethical imperatives that must guide each step on the path towards greater social justice and comprehensive sustainability, comprehensive and enduring over time. The collective commitment to these principles is undoubtedly the key to opening the doors to a more promising future for all.

## 1.1 Democracy and Transparency:
   Cornerstones of Comprehensive Multidimensional Social Wellbeing

In contemporary politics, democracy stands as a political system ardently seeking to ensure both citizen participation and the staunch protection of their rights.[5] An enduring democracy fundamentally necessitates transparency in public administration, providing citizens the mechanism to exert significant oversight over their governing bodies. This transparency is not merely a suggestion but a quint essential principle for robust democratic governance. It grants citizens access to crucial information regarding public management, thereby bolstering both accountability and civic participation. It signifies an open book policy where governmental actions are not shrouded in secrecy but laid bare for public scrutiny and assessment. See Figure 1.

Figure 1
The Triumvirate of Democracy, Transparency and
Social Wellbeing in Contemporary Politics

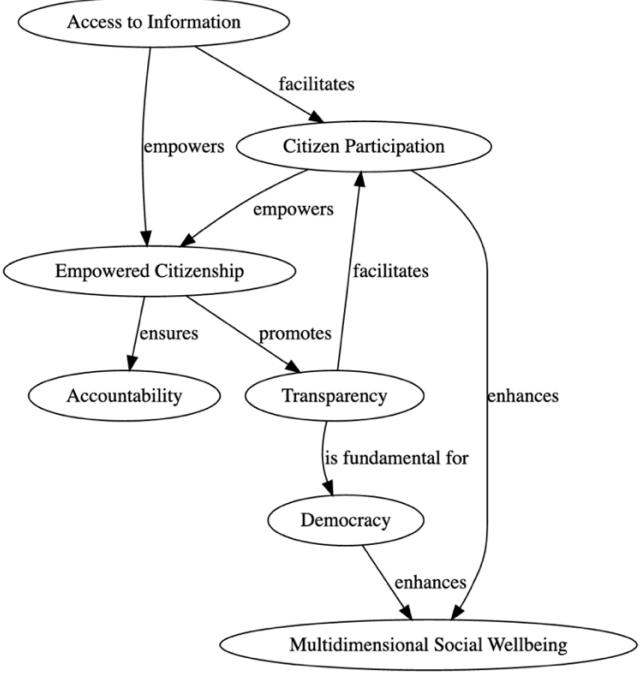

Source: Own elaboration

---

[5] (Cienfuegos, 2023) highlighted that in Latin America, democratic regimes have failed to alleviate poverty and address the needs of the population. This region stands out as the world's poorest in terms of the quality of its democracies. See. Cienfuegos Terrón, Marco Aurelio. La democracia como factor de desigualdad en América Latina y México. Quivera Revista de Estudios Territoriales, [S.l.], v. 25, n. 2, p. 9-31, jun. 2023. ISSN 2594-102X. https://quivera.uaemex.mx/article/view/20400. Fecha de acceso: 06 oct. 2023 doi: https://doi.org/10.36677/qret.v25i2.20400.

In tandem with democracy and transparency, the multidimensional social wellbeing,[6] emerges as another pivotal aspect, referring to the living conditions of citizens across various domains such as health, education, housing, among others. A democratic state ardently endeavors to advocate and enforce public policies aimed at ensuring adequate levels of social wellbeing. This triumvirate of democracy, transparency and social wellbeing is intertwined in a relationship of mutual reinforcement, where each element bolsters the other in achieving a society that upholds the rights[7] and wellbeing of its citizens. Democracy hinges on transparency for citizens to appraise and evaluate public policies[8] directed towards their social wellbeing. Concurrently, social wellbeing is significantly enhanced by the democratic participation of citizens, allowing their voices and concerns to steer the policies affecting their lives.

However, this idyllic vision is often mired in challenges, particularly in regions like Latin America, where the frailty of democracies is ostensibly tied to low levels of transparency[9] and substantial difficulties in ensuring the social welfare of extensive population sectors. Scholars postulate this intricate connection, observing the palpable issues stemming from the lack of transparency,[10] leading to diminished welfare conditions and a weakened democratic fabric.

The limitations faced by citizens in acquiring crucial information regarding public management, policy formations and the deployment and application of public assets stifles their meaningful involvement in the democratic pathway.[11] This blockage culminates in a burgeoning sense of disenchantment, estrangement, and doubt regarding the democratic structure.

Considering these issues, revitalizing democracy in such environments demands the immediate introduction of sweeping changes. These reforms should ardently advocate for enhanced transparency concerning the handling of public funds and assure the seamless and meaningful engagement of the populace in welfare policy discussions and implementations. The enhancement of transparent practices within governmental operations and ensuring citizens' engagement will lay the foundation for a more robust and resilient democratic environment, fostering trust and collaboration between the governed and the governing entities.[12]

An empowered citizenry, armed with access to information, becomes a formidable force in ensuring accountability, promoting transparent governance actively participating in the democratic process to safeguard and enhance their welfare conditions.[13]

---

[6] See. Arocena, F.A., Aguilar, H.S., Madrigal, Y.O., & Ceballos, J.C. (2011). Percepciones de bienestar social, anomia, interés e impotencia política en relación con las actitudes hacia la democracia. Liberabit. Revista de Psicología, vol. 17, núm. 1, enero-junio, 2011, pp. 7-17.

[7] See. Fernández, R., & Angel, M.A. (2014). Educación y participación ciudadana en la democratización de la Administración Local: realidades y perspectivas de futuro en Galicia. Universidade de Santiago de Compostela Facultad de Ciencias da Educación. Departamento de Teoría da Educación, Historia da Educación e Pedagoxía Social. Tesis doctoral. https://minerva.usc.es/xmlui/bitstream/10347/12004/1/rep_751.pdf

[8] See. Castillo Liendo, Y.D. (2023). Control fiscal externo desde la complementariedad del control ciudadano de las políticas públicas. Revista de Ciencias Sociales. Revista De Ciencias Sociales, (177), 123–134. https://doi.org/10.15517/rcs.v0i177.54041

[9] See. Christofoletti, R., & Becker, D. (2020). Retos para la adopción de la transparencia en la agenda de ética periodística en América Latina. *Sintaxis, 1(5), 11–30. https://doi.org/10.36105/stx.2020n5.01*

[10] See. Valencia Escamilla, L. (2016). Rendición de cuentas y los mecanismos de transparencia legislativa en América Latina / Rendered Accounting and mechanisms of Legislative Transparency in Latin America. RICSH Revista Iberoamericana De Las Ciencias Sociales Y Humanísticas, 5(10), 49 - 74. Recuperado a partir de https://www.ricsh.org.mx/index.php/RICSH/article/view/76

[11] See. Bianchetto, A. (2022). América Latina entre rebelión y estado de excepción permanente. Revista revoluciones. -123- Vol. 4, N° 8 (2022), pp. 122-135. http://revistarevoluciones.com/index.php/rr/article/view/85

[12] See. A. Nigmatov, A. Pradeep and N. Musulmonova, "Blockchain Technology in Improving Transparency and Efficiency in Government Operations," 2023 15th International Conference on Electronics, Computers and Artificial Intelligence (ECAI), Bucharest, Romania, 2023, pp. 01-06, doi: 10.1109/ECAI58194.2023.10194154

[13] See. Negash, Y. T., Sarmiento, L. S. C., Tseng, S. W., Lim, M. K., & Tseng, M. L. (2023). A circular waste bioeconomy development model in the Ecuadorian fishery industry: the impact of government strategy on supply chain integration and smart operations. Environmental science and pollution research international, 30(43), 98156–98182. https://doi.org/10.1007/s11356-023-29333-8

The evident link between democracy, transparency and social welfare holds the key to strengthening democratic systems, fostering transparency, and enhancing the welfare conditions of citizens, ensuring a society where the rights and wellbeing of every citizen are upheld and protected.[14]

**1.1.1 Trust and credibility: The paramountcy of enhanced transparency**

In the realm of governance and public administration, trust and credibility are essential currencies. Enhanced transparency stands as the linchpin for nurturing and bolstering these vital elements. This subsection delves into the paramount importance of transparency in building trust and credibility within governments, organizations, and societies.[15] By shedding light on the profound impact of transparent communication, ethical accountability, and open dialogue, we explore how transparency serves as the foundation upon which trust is built and credibility is earned. In an era marked by information flow, understanding the role of transparency is crucial in maintaining public faith and fostering lasting partnerships.[16]

**1.1.2 Establishing trust through augmented Transparency**

Transparency stands as a pivotal instrument for instilling trust, elevating credibility, and refining the ethical stance of organizations across diverse sectors.[17] It amplifies the assurance of citizens, employees and various stakeholders in both governments and organizations by guaranteeing the swift and accurate release of vital information. This element gains prominence during times of crisis such as epidemics, wherein transparent, straightforward, and sincere communication can augment the credibility of governments and reinforce public trust.[18]

In today's digital world, transparency acts as a catalyst for open communication, a critical component for trust-building.[19] Platforms like social media and other digital tools function as mediums for transparent dialogue between governments, organizations, and citizens, enhancing understanding and fostering the exchange of insights and perspectives.[20] This interactive conversation empowers citizens to express their concerns, seek information and offer feedback, thus augmenting mutual understanding and collaboration.

**1.1.3 Fostering accountability through Transparency**

Transparency also significantly enhances social and ethical accountability among governments and organizations. The public disclosure of information allows various stakeholders, including the public and media, to oversee organizational and governmental operations, ensuring responsibility and ethical behavior.[21] Awareness of their activities being subject to public scrutiny propels organizations towards ethical actions and a focus on enduring, sustainable objectives. It spurs them to assess the extensive impacts of their operations and make decisions that positively resonate with society and the environment.

---

[14] See. Medel-Ramírez, Carlos, Medel-López, Hilario and Lara-Mérida Jennifer (2023) Digital Governance in the 21stCentury: The LiTCoDE Framework for Transparency, Leadership, and Technological Evolution a Comparative Study of Mexico and Vietnam. September 2023. DOI: 10.13140/RG.2.2.33060.24963 License CC BY 4.0.

[15] See. Cervantes Hernández, A. (2023). Gobierno abierto y transparencia gubernamental: una perspectiva desde los gobiernos municipales de México. Encrucijada Revista electrónica Del Centro De Estudios En Administración Pública, (43), 1–16. https://doi.org/10.22201/fcpys.20071949e.2023.43.84391

[16] See. Armeaga García, F., & Medrano González, R. (2023). Testigos sociales y rendición de cuentas. Análisis de la experiencia en el Estado de México. Dilemas contemporáneos: Educación, Política y Valores. Año: X Número: 3. Artículo no.:83 Período: 1ro de mayo al 31 de agosto del 2023. https://dilemascontemporaneoseducacionpoliticayvalores.com/index.php/dilemas/article/view/3669/3613

[17] See. Tiwari, A.K., & Deshpande, A. (2022). Relationship Transparency-Trust Cycle: A Matter of Trust and Competency for Frontline Managers. Cardiometry. Issue 23; August 2022; p.476-488; DOI: 10.18137/cardiometry.2022.23.476488; https://www.cardiometry.net/issues/no23-august-2022/relationship-transparency-trust-cycle

[18] See. Rieznik, S., & Lee, H. (2021). Citizens' Perception of Corruption and Transparency as Determinants of Public Trust in Local Government in Ukraine. Hrvatska i komparativna javna uprava. Hrvatska i komparativna javna uprava: časopis za teoriju i praksu javne uprave, Vol. 21 No. 2, 2021. https://hrcak.srce.hr/file/378568

[19] See. Husni, M., Damayanti, R.A., & Indrijawati, A. (2023). The role of the village government performance and transparency in influencing village public trust. Journal of Accounting and Investment. https://journal.umy.ac.id/index.php/ai/article/view/17114/pdf

[20] See. Hartanto, D., & Siregar, S.M. (2021). Determinants of Overall Public Trust in Local Government: Meditation of Government Response to COVID-19 in Indonesian Context. Transforming Government: People, Process and Policy. Transforming Government: People, Process and Policy, Vol. 15 No. 2, pp. 261-274. https://doi.org/10.1108/TG-08-2020-0193

**1.1.4 Amplifying societal connections through Transparency**

Beyond the realms of trust and accountability, transparency bolsters societal connections by nurturing mutual understanding, boosting participation, and advocating collective values. The distribution of transparent information cultivates mutual understanding among governments, businesses, and citizens, fostering unity and collaboration. It empowers citizens with the knowledge to offer valuable feedback on governmental policies and services, ensuring their voices resonate and their concerns find solutions.

Societies imbibing values of openness, tolerance and equality tend to harbor governments and organizations that function transparently, reflecting a dedication to these principles and contributing to societal harmony and unity.[22] The transparency emerges as a robust tool for establishing trust, nurturing accountability, amplifying societal connections. By enhancing perceptions, guaranteeing accountability, and championing shared values, it unites individuals and enriches the relationships between governments, organizations, and citizens.[23]

In an epoch where information holds paramount significance, the embrace and advancement of heightened transparency by organizations and governments yield boundless benefits. This devotion to openness and accountability will ultimately culminate in the construction of a more trustful, accountable, and united society.

## 1.2 Effective governance: Exploring the models of Finland, Singapore, and New Zealand

The exemplary governance models of Finland, Singapore and New Zealand stand as lighthouses in the global spectrum, each exhibiting strengths, proven strategies consistent policies that yield high standards of living and robust economies.

Finland, nestled in the Northern reaches of Europe, routinely shines in global assessments related to governance. It's not just the icy glimmer of its vast expanses of snow and forest that attracts global admiration, but its unwavering commitment to law, transparency, and effective governance. The nation is perpetually at the forefront when it comes to minimal corruption,[24] transparent and accountable operations and a robust rule of law upheld by an independent judiciary[25]. Its governance model centers on a highly professional and autonomous civil service, rooted in consensus-oriented policymaking that embraces the voices and concerns of a myriad of stakeholders.[26] This inclusive approach has fostered an advanced welfare state, where equality isn't just aspired to but actively cultivated, resulting in impressive quality of life indicators that serve as a global benchmark.

---

[21] See. Qi Zheng (2023) Restoring trust through transparency: Examining the effects of transparency strategies on police crisis communication in Mainland China. Public Relations Review, Volume 49, Issue 2, 2023. https://doi.org/10.1016/j.pubrev.2023.102296

[22] See. Kwan, D., Cysneiros, L.M., & Leite, J.S. (2021). Towards Achieving Trust Through Transparency and Ethics. 2021 IEEE 29th International Requirements Engineering Conference (RE), 82-93. https://arxiv.org/abs/2107.02959

[23] See. Home Arias, P., & Arévalo, J. C. (2021). La transparencia y la rendición de cuentas mecanismos del "gobierno abierto" como instrumento de compromiso público y responsabilidad democrática en las organizaciones públicas. Documentos de Trabajo. ECACEN, 1. https://doi.org/10.22490/ECACEN.4693

[24] See. Minna Kimpimäki (2018) Corruption in a non-corrupt country: what does corruption look like in Finland?, International Journal of Comparative and Applied Criminal Justice, 42:2-3, 233-252, DOI: 10.1080/01924036.2017.1310662

[25] See. Lidén, K. (2023). A better foundation for national security? The ethics of national risk assessments in the Nordic region. Cooperation and Conflict, 58(1), 3-22. https://doi.org/10.1177/00108367211068877

[26] See. O. C. Osifo, "A Study of Coordination Challenges in Digital Policy Implementation and Evaluation in Finland," 2020 43rd International Convention on Information, Communication and Electronic Technology (MIPRO), Opatija, Croatia, 2020, pp. 1402-1409, doi: 10.23919/MIPRO48935.2020.9245438. https://osuva.uwasa.fi/bitstream/handle/10024/11810/Osuva_Osifo_2020.pdf

A continent away, the city-state of Singapore narrates a different, yet equally compelling story of governance excellence[27]. Despite its small geographical size, Singapore's impact on global governance standards is colossal.[28] It has spectacularly transformed from a fledgling nation grappling with a myriad of challenges to a high-income powerhouse,[29] synonymous with effective, visionary leadership and a lean, efficient, and meritocratic bureaucracy. See Figure 2.

Figure 2
Governance Models Comparison: Finland, Singapore and New Zealand

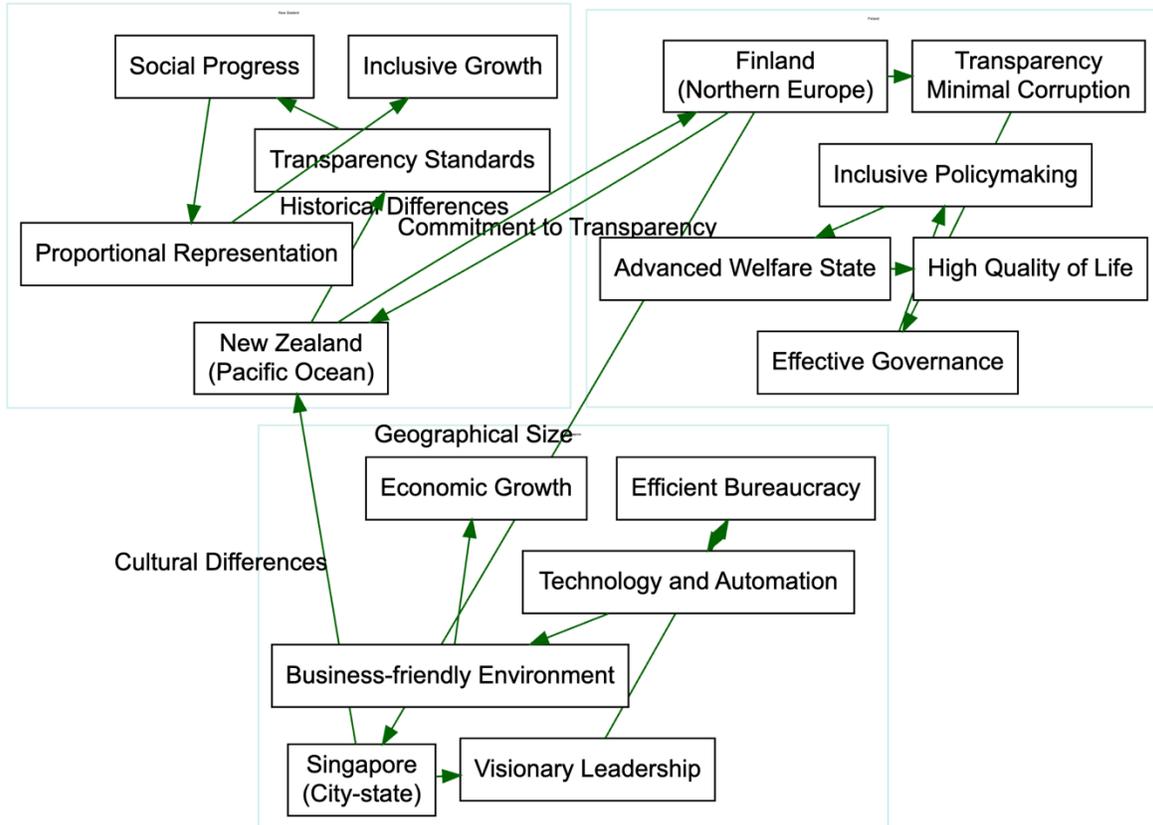

Source: Own elaboration

This transformation has been bolstered by high public sector wages,[30] which serve as a magnetic pull for talent while concurrently minimizing corruption.[31] The heavy incorporation of technology and automation within its public services underscores Singapore's commitment to efficiency and innovation, fostering a business-friendly environment that propels economic growth and technological advancement.

---

[27] See. Ng, C.J.W. (2019). Governing (Through) Affect: A Social Semiotic Perspective of Affective Governance in Singapore. In: Rajandran, K., Abdul Manan, S. (eds) Discourses of Southeast Asia. The M.A.K. Halliday Library Functional Linguistics Series. Springer, Singapore. https://doi.org/10.1007/978-981-13-9883-4_2

[28] See. Master, D.C., Ngoc Huy, D.T., Huong Sang, N., & Thin, V.D. (2022). Enhancing Vietnam Bank Management and Governance via References of Several Northern Asian Corporate Governance Standards after the Global Crisis - Cases in Singapore and Pakistan. Asian Journal of Applied Science and Technology. Volume 6, Issue 2, Pages 49-63, April-June 2022

[29] See. Menon, S.V. (2007). Governance, leadership, and economic growth in Singapore. MPRA Paper No. 4741, posted 06 Sep 2007 UTC. https://mpra.ub.uni-muenchen.de/4741/1/MPRA_paper_4741.pdf

[30] See. Guan, L.S. (1997). Sustaining excellence in government: the Singapore experience. Public Administration and Development, Volume 17, Issue 1. February 1997. 167-174.

[31] See. Le Queux, S. and Kuah, A.T.H. (2021), "君子 Junzi leadership in Singapore: governance and human capital development", Journal of Management Development, Vol. 40 No. 5, pp. 389-403. https://doi.org/10.1108/JMD-05-2019-0194

New Zealand, enveloped by the vast Pacific Ocean, stands as another beacon of governance excellence.[32] The island nation is globally recognized for its robust transparency standards,[33] minimal corruption,[34] and consistent social progress. It robustly embraces a proportional representation electoral system, ensuring a broad, equitable representation that translates into governance marked by high effectiveness and regulatory quality. The New Zealand governance model,[35] characterized by an independent and non-partisan public service, unyieldingly focuses on inclusive growth, continually seeking to eradicate corruption and enhance public services.[36]

In the grand tapestry of global governance, the threads of Finland,[37] Singapore and New Zealand intertwine to form a compelling picture of what's attainable. Despite their geographical, cultural, and historical differences, these nations share a common thread: a steadfast commitment to transparent institutions, meritocratic bureaucracies, and a categorical rejection of corruption. In these lands, inclusive policymaking isn't just a theoretical concept discussed in hallowed halls but a living, breathing reality, consistently fueling progress, prosperity, and the well-being of their citizens.

In examining the governance models of Finland,[38] Singapore[39] and New Zealand,[40] countries around the world can unearth valuable insights and practical strategies to bolster their own governance frameworks. From Finland's consensus-oriented policymaking to Singapore's visionary leadership and New Zealand's focus on inclusive growth and transparency, these nations offer a treasure trove of lessons for global leaders seeking not just to improve governance, but to elevate the lives and well-being of their citizens, ultimately contributing to global progress, stability, and peace.

## 1.3 Democracy unveiled: Challenges, insights, and ethics

The pillars of a robust democratic society, one that stands resilient in the face of myriad challenges, are intricately tied to transparency,[41] social justice, and sustainability. The complex relationship between democracy and various hurdles, including ecological crises in capitalist societies, the dearth of female representation in political leadership and the need for innovation in leadership recruitment and governance, outlines the critical landscape where democracy is tested.

---

[32] See. Gregory, R. and Zirker, D. (2013), "Clean and Green with Deepening Shadows? A Non-Complacent View of Corruption in New Zealand", Different Paths to Curbing Corruption (Research in Public Policy Analysis and Management, Vol. 23), Emerald Group Publishing Limited, Bingley, pp. 109-136. https://doi.org/10.1108/S0732-1317(2013)0000023005

[33] See. Edeigba, Jude and Amenkhienan, Felix (2017) The Influence of IFRS Adoption on Corporate Transparency and Accountability: Evidence from New Zealand, Australasian Accounting, Business and Finance Journal, 11(3), 2017, 3-19. doi:10.14453/aabfj.v11i3.2

[34] Zirker, D. (2017), "Success in combating corruption in New Zealand", Asian Education and Development Studies, Vol. 6 No. 3, pp. 238-248. https://doi.org/10.1108/AEDS-03-2017-0024

[35] See. Matthewman, S. (2017). 'Look no Further than the Exterior': Corruption in New Zealand. https://apo.org.au/sites/default/files/resource-files/2017-11/apo-nid129671.pdf

[36] See. Scott, R. J., Donadelli, F., & Merton, E. R. (2022). Administrative philosophies in the discourse and decisions of the New Zealand public service: is post-New Public Management still a myth? International Review of Administrative Sciences, 0(0). https://doi.org/10.1177/00208523221101727

[37] See. Gregory, R., & Zirker, D.R. (2022). Historical corruption in a 'non-corrupt' society: Aotearoa New Zealand. Public Administration and Policy. Public Administration and Policy. Vol. 25 No. 2, 2022 pp. 150-162. 1727-2645. DOI 10.1108/PAP-01-2022-0008

[38] See. Gregory, R. and Zirker, D. (2022), "Historical corruption in a 'non-corrupt' society: Aotearoa New Zealand", Public Administration and Policy: An Asia-Pacific Journal, Vol. 25 No. 2, pp. 150-162. https://doi.org/10.1108/PAP-01-2022-0008

[39] See. Mamani Mamani Enrique Jotadelo et al. (2022). El Singapur, un modelo de innovación educativa a seguir para la transformación económica y social. Waynarroque - Revista de ciencias sociales aplicadas. https://unaj.edu.pe/revistacientificawaynarroque/index.php/rcsaw/article/view/43/35

[40] See. Christensen, T., & Lægreid, P. (2014). La nueva administración pública: el equilibrio entre la gobernanza política y la autonomía administrativa. Revista Do Serviço Público, 52(2), p. 68-109. https://doi.org/10.21874/rsp.v52i2.306

[41] See. Bazan Cruz, M. (2023). Transparentar: una acción polimórfica y estratégica en democracia. Revista de Ciencias Sociales. Rev. Ciencias Sociales 181: 55-70 / 2023 (III)ISSN Impreso: 0482-5276. https://revistas.ucr.ac.cr/index.php/sociales/article/view/56702/57318

### 1.3.1 Challenges

One of the significant challenges is the interaction between democracy and ecological crises[42] in capitalist societies. Capitalism's relentless pursuit of growth and profit often clashes with environmental preservation and sustainability. Notwithstanding the democratic framework, the accumulation of power and wealth within capitalist systems often erodes the collective determination to tackle ecological challenges[43] effectively.

Another pressing issue pertains to the inadequate representation of women[44] in political leadership roles[45] This shortfall dilutes the democratic essence, marginalizing a substantial section of the population from decision-making processes. The significant lack of diverse perspectives and experiences hinders the development of holistic and inclusive policies and strategies.

Furthermore, existing approaches to leadership recruitment and governance are becoming increasingly inadequate in navigating the complex,[46] interconnected challenges of our time. Traditional models often lack the flexibility, inclusiveness, and innovative outlook crucial for addressing contemporary issues. In the document, these intersections between information,[47] uncertainty, transparency, democracy, and social wellbeing are thoroughly examined and analyzed.

### 1.3.2 Insights

Amid these challenges, valuable insights emerge that light the path forward. The importance of reimagining education as a tool for democracy stands paramount. Education tailored for democratic enrichment fosters informed, engaged and critical citizens, pivotal for the sustainable growth and evolution of democratic societies.[48] Such an approach in education transcends traditional learning paradigms, embedding civic responsibility and democratic values at its core. See Figure 3.

Figure 3

Pursuit of Social Wellbeing

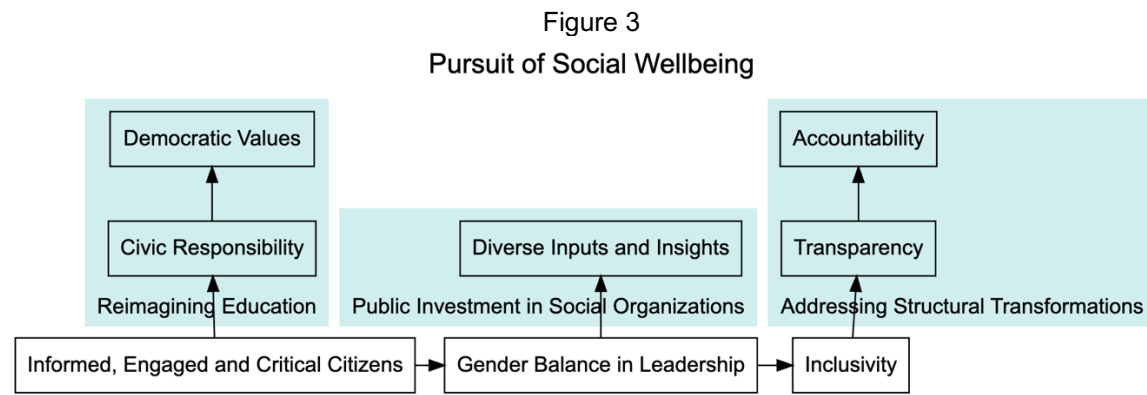

Source: Own elaboration

---

[42] See. Gonzales-Loli, M., Sanabria-Boudri, F., Ríos-Garay, J., & Colina-Ysea, F. (2021). Crecimiento económico y políticas ambientales en Latinoamérica. CIENCIAMATRIA, 7(1), 14-34. https://doi.org/10.35381/cm.v7i1.461
[43] See. Fidélis, T., Teles, F., Roebeling, P., & Riazi, F. (2019). Governance for Sustainability of Estuarine Areas Assessing Alternative Models Using the Case of Ria de Aveiro, Portugal. Water.
[44] See. Rivas Maldonado, Fan Jua. 2022. "Financiamiento público que promueve el liderazgo político de las mujeres: hallazgos a partir de la fiscalización realizada por el INE de México". Elecciones (julio-diciembre), 21(24): 99-124. DOI:10.53557/Elecciones.2022.v21n24.03
[45] See. García Méndez, E. (2019). Representación política de las mujeres en los Congresos subnacionales en México. Un modelo de evaluación. Estudios Políticos, Num. (46). https://doi.org/10.22201/fcpys.24484903e.2019.46.68289
[46] See. Silvestri, M., Tong, S., & Brown, J. (2013). Gender and Police Leadership: Time for a Paradigm Shift? International Journal of Police Science & Management, 15(1), 61-73. https://doi.org/10.1350/ijps.2013.15.1.303
[47] See. Navarra, D.D., & Cornford, T. (2012). The State and Democracy After New Public Management: Exploring Alternative Models of E-Governance. The Information Society, 28, 37 - 45.DOI: 10.1080/01972243.2012.632264
[48] See. Olawale, B.E., Mncube, V., & Harber, C. (2023). Popular conceptions of democracy in a mathematics teacher-education programme. South African Journal of Education.

Moreover, public investment in social organizations emerges as a significant insight. Bolstering institutions that champion women's rights and representation not only corrects the gender imbalance in leadership roles but also enriches the democratic fabric with diverse, multi-faceted inputs and insights. Further, addressing structural transformations in the public arena is indispensable.[49] Structural shifts that prioritize inclusivity, transparency and accountability enhance the democratic framework, fostering a more resilient and robust political and social structure.

### 1.3.3 Ethical imperatives

At the heart of the pursuit for a robust democratic society lies a moral and ethical commitment. Democracy must be inclusive, pluralistic, and respectful of all individuals and cultures.[50] The ethical imperative mandates the dismantling of the challenges faced by capitalist and fossil-fuel-based democracies.[51] It calls for the nurturing of democratic values and practices through active civic engagement and the instillation of a culture of ethical business conduct and accountability in corporate governance.[52] See Figure 4.

Figure 4
Exploration Axioms Democracy, Transparency and Social Wellbeing

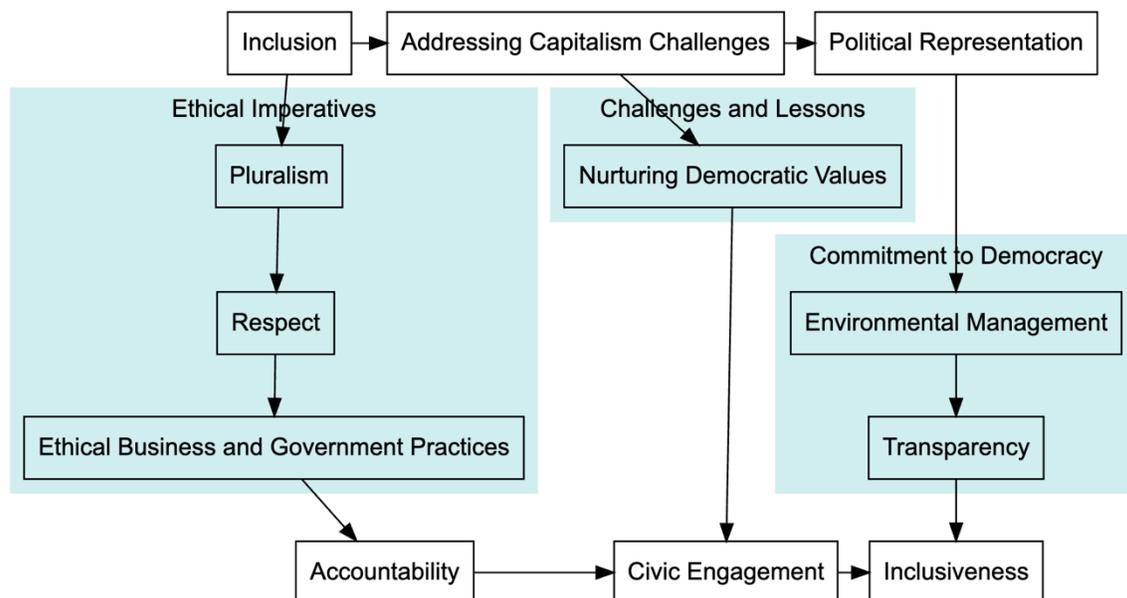

Source: Own elaboration

In the labyrinth of challenges, the lessons learned, and the ethical imperatives identified provide a beacon for reinforcing democratic strength.

---

[49] See. Carmona Gallego, D. (2021). El duelo en el ámbito público: Composiciones en torno a la ética del cuidado, la ontología de la vulnerabilidad y la interdependencia. Del Prudente Saber Y El máximo Posible De Sabor, (14), 22–35. https://doi.org/10.33255/26184141/1112
[50] See. Lafferty, W.M., Langhelle, O. (1999). Sustainable Development as Concept and Norm. In: Lafferty, W.M., Langhelle, O. (eds) Towards Sustainable Development. Palgrave Macmillan, London. https://doi.org/10.1057/9780230378797_1
[51] See. Oleinykov, S. (2022). Ethics and legal aspects of public institutions' legal activities. Grail of Science. https://archive.journal-grail.science/index.php/2710-3056/article/view/173
[52] See. Pérez, J.R., & Canizales, R.R. (2021). Estado de derecho y transparencia : Un acercamiento desde la historia, el derecho y la ética. Misión Jurídica. Revista Misión Jurídica / ISSN 1794-600X / E-ISSN 2661-9067. Vol. 14 - Número 20 / Enero - Junio de 2021 / pp. 70 - 84

The collective endeavor must focus on surmounting the obstacles, embracing the insights, and adhering to the ethical obligations in the pursuit of a more transparent, inclusive, and accountable society. The commitment to addressing issues from political representation to environmental management stands as a testament to the unwavering resolve to fortify democratic structures. The journey, though arduous, is vital for achieving lasting social justice and sustainability, ensuring the thriving of diverse communities within the robust bastions of democracy. See Figure 5

Figure 5
Key Interconnections in the Pursuit of Social Wellbeing

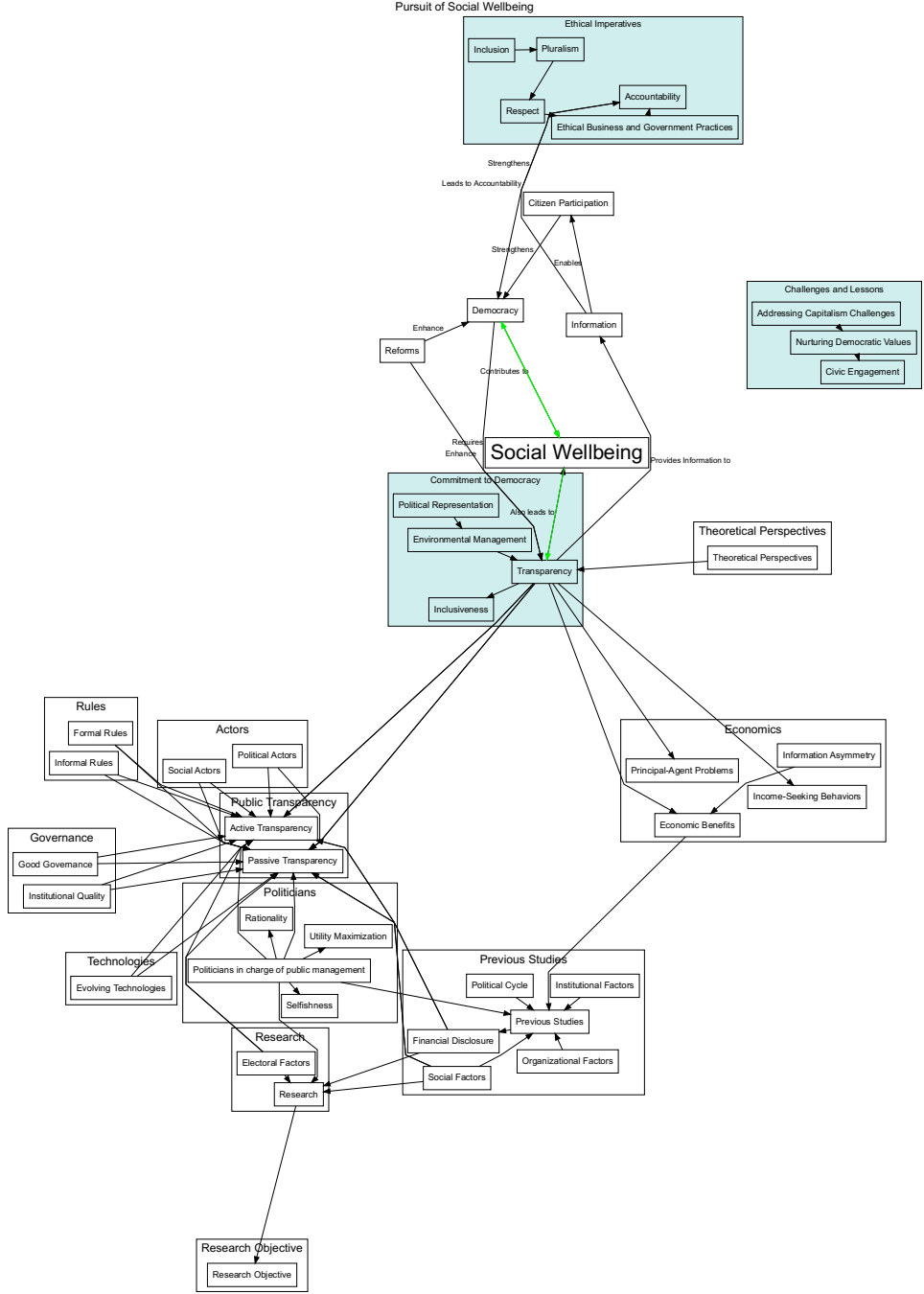

Source: Own elaboration

According to Figure 5, the exploration framework has unveiled the intricate and interdependent nature of the core principles that underpin our society. From the unwavering pursuit of social wellbeing to the dynamic interplay of democracy, transparency, and ethical imperatives, it becomes clear that our world operates within a multifaceted tapestry of relationships and influences that define our progress and collective ambitions.

At its core, democracy stands as a guiding beacon, empowering individuals to actively participate in shaping the trajectory of their shared destiny. Its vitality is magnified when paired with transparency, a force that ensures the unobstructed flow of information and holds those in positions of authority accountable for their decisions. Accountability, in turn, safeguards the integrity of democracy, nurturing the trust of the public in their governing institutions.

Social wellbeing emerges as the goal, encapsulating the holistic welfare and prosperity of society. It underscores the paramount significance of policies and systems designed to prioritize the diverse needs and aspirations of individuals, thus fostering a sense of unity and shared purpose.

In this intricate web of concepts, ethical imperatives, including inclusivity, pluralism, respect, and ethical practices, serve as our moral compass. They guide us through the labyrinth of governance, reminding us of our enduring commitment to fairness, equity, and integrity in society.

Challenges are an inevitable part of our journey, and the ongoing need to cultivate democratic values and civic engagement remains ever-present. Recognizing potential pitfalls, particularly within systems like capitalism, is essential. Continually working to address these challenges while steadfastly upholding the core tenets of democracy is imperative.

Our commitment to democracy extends far beyond the realms of politics, encompassing environmental stewardship, inclusiveness, and transparency in every facet of society. These facets collectively fortify our dedication to democratic ideals as we navigate the complexities of the modern world.

In our quest for a comprehensive understanding of these concepts, we have delved into additional subgraphs, each offering unique insights into the intricate tapestry of governance. These subgraphs illuminate the nuances of governance, rules, technology, economics, and more, enriching our comprehension of our societal landscape.

This exploration serves as a poignant reminder that the progress and wellbeing of our society hinge upon the interconnected interplay of democratic principles, transparency, accountability, and ethical conduct. It is an impassioned call to action, not solely for policymakers and scholars but for every individual who shares the vision of a more equitable, informed, and flourishing world.

The unwavering commitment to addressing these challenges, embracing insights, and adhering to ethical imperatives is pivotal. The fortification of democratic structures, the advancement of transparency, inclusiveness, and accountability across various dimensions of society, from politics and education to corporate governance and environmental management, is imperative for the flourishing of democracy. It is a journey essential for achieving lasting social justice, sustainability, and the thriving of diverse communities within the robust realms of democratic societies.

## 1.4 Preliminary conclusions

Our deep dive into Chapter 1, has illuminated the paramount significance of transparency and democracy in our unwavering quest to build societies that are fairer, more equitable and sustainable. These twin pillars aren't mere abstractions; they represent ethical imperatives that form the bedrock of a democracy that is both sturdy and adaptable.

They are the guiding principles upon which we can construct societies that prioritize the holistic wellbeing of every individual. Transparency emerges as a potent force, nurturing trust, fostering accountability, and weaving a resilient social fabric.

Our examination of Finland, Singapore and New Zealand's exemplary governance models has bestowed upon us invaluable insights into the transformative power of transparent and accountable governance. These nations stand as luminous beacons, demonstrating how the unwavering commitment to democratic principles and transparency can pave the way for high standards of living, economic prosperity, and the overall betterment of citizens. Their success stories underscore the tangible benefits of upholding these principles consistently.

We have also grappled with the formidable challenges that modern democracies face, from the clash between capitalism and ecological sustainability to the glaring underrepresentation of women in leadership roles. Yet, within these challenges, we have unearthed vital lessons. Education emerges as a potent tool for nurturing informed and engaged citizens, while the support for women's organizations becomes a catalyst for achieving gender equality in leadership.

In conclusion, our journey through this chapter serves as a stark reminder of our ethical duty to uphold transparency and democracy. These principles are not mere aspirations; they are the moral compass guiding us toward societies that cherish social justice, inclusivity, and sustainability. This journey demands unwavering dedication, adaptability and a collective effort encompassing governments, civil society, the private sector, and every individual.

# Chapter 2
# An Inferential Framework for Democracy, Transparency and Social Wellbeing: Key Theorems and Fundamental Axioms

In an era characterized by a rapid surge in technological advancements and the ever-evolving socio-political landscape, the very foundations of democracy,[53] transparency, and social wellbeing, find themselves at the precipice of profound change.[54] In navigating the intricate complexities of our increasingly globalized world, the pressing need for a rigorous inferential framework to steer the course of policymaking and governance becomes undeniably conspicuous. This chapter embarks on a journey through a complex tapestry of concepts, theorems and axioms that constitute the bedrock of an inferential framework meticulously crafted to elevate democracy, transparency, and social wellbeing to new heights.[55]

Transparency, conversely, assumes a central role in the mechanics of democratic systems.[56],[57] This chapter aspires to dissect the intricate symbiosis between democracy and transparency, bringing to light pivotal theorems that illuminate the path toward a more accountable and participatory form of governance. Social wellbeing, an intricate and multi-dimensional construct, enshrouds the composite quality of life and welfare of a society's citizenry.[58] It encapsulates the collective prosperity, health, educational attainment, and overall contentment of a populace.[59] Within the framework of inferential reasoning aimed at fostering democracy and transparency, social wellbeing assumes the dual role of both a guiding beacon and an evaluative gauge of achievement. Through the establishment of fundamental axioms that prioritize the augmentation of social wellbeing, policymakers find the compass with which to navigate their decisions, all in service to the holistic betterment of society.

---

[53] See. Naomi Roht-Arriaza, Democracia y transparencia en el SIDH: una experiencia en marcha, 8 Rev. Direito & Práxis. 1652 (2017). DOI: 10.12957/dep.2017.28036| ISSN: 2179-8966 https://repository.uchastings.edu/faculty_scholarship/1568

[54] See. Zhironkin, S., Gasanov, M., & Zhironkina, O. (2016). The Analysis of Social Wellbeing Indicators in the Context of Russian Economy Structural Changes. In F. Casati (Ed.), Lifelong Wellbeing in the World - WELLSO 2015, vol 7. European Proceedings of Social and Behavioural Sciences (pp. 124-131). Future Academy. https://doi.org/10.15405/epsbs.2016.02.17

[55] See. Arsenault, Amelia C., and Sarah E. Kreps, 'AI and International Politics', in Justin B. Bullock, and others (eds), The Oxford Handbook of AI Governance (Online Ed, Oxford Academic, 14 Feb. 2022), https://doi.org/10.1093/oxfordhb/9780197579329.013.49

[56] See. Finol-Romero, L.T. (2019). Transparencia, corrupción y democracia en América Latina. teoría y praxis. Revista Espacios. Vol. 40 (Nº 27) Año 2019. Pág. 17

[57] See. Abad Alcalá, L. (2023). Transparencia y rendición de cuentas ante la crisis de legitimidad del Estado democrático. Revista Española de la Transparencia. número 16 (Primer semestre. Enero - junio 2023) https://revistatransparencia.com/ojs/index.php/ret/article/view/272/347

[58] See. Acosta-Rosero, D. (2022). La importancia del estado de bienestar en la economía social y solidaria: Eslabón democrático y herramienta de transición. Revista Nacional De Administración, 13(2), e4480. https://doi.org/10.22458/rna.v13i2.4480

[59] See. Quiroga Juárez, C. A., & Villafuerte Valdés, L. F. (2023). Estudio del Barómetro de las Américas en un marco de convergencia de la cohesión social con el desarrollo y bienestar: caso México. Revista Mexicana De Opinión Pública, (34). https://doi.org/10.22201/fcpys.24484911e.2023.34.82880

## 2.1 The Mathematical Framework: Interconnections Between Democracy, Transparency and Social Wellbeing

To forge an inferential framework capable of synthesizing these three pivotal pillars—democracy, transparency, and social wellbeing—requires a deep dive into an intricately interconnected web of theorems and foundational axioms.

These principles collectively provide a structured framework for navigating the intricate terrain of decision-making and policy development, transcending the boundaries of time and place that frequently constrain our vision.[60] In essence, they act as the intellectual scaffolding upon which we can construct a society that is not only more just, transparent, and responsive but also one that wholeheartedly embodies the core principles of democracy, transparency, and the holistic social wellbeing of its entire populace.[61]

The complex relationships between democracy, transparency and social wellbeing can be represented symbolically and mathematically. This can be in the form of an inferential model using symbols to represent these key concepts and their interrelations.

**Notation:**

Let:

- D     represent Democracy
- T     represent Transparency
- SW   represent Social Wellbeing
- R     represent Reforms
- I     represent Information
- A     represent Accountability
- C     represent Citizen Participation

**Relationships:**

1. $D \rightarrow T$: Democracy inherently requires Transparency.
2. $T \rightarrow I$: Transparency provides Information to citizens.
3. $I \rightarrow A$: Information leads to Accountability.
4. $I \rightarrow C$: Information enables Citizen Participation.
5. $A, C \rightarrow D$: Accountability and Citizen Participation strengthen Democracy.
6. $D \rightarrow SW$: Democracy contributes to enhanced Social Wellbeing.
7. $T \rightarrow SW$: Transparency also leads directly to enhanced Social Wellbeing.
8. $R \rightarrow (D, T)$: Reforms enhance both Democracy and Transparency.
9. $D \rightarrow T$: Democracy inherently requires Transparency.
10. $T \rightarrow I$: Transparency provides Information to citizens.
11. $I \rightarrow A$: Information leads to Accountability.
12. $I \rightarrow C$: Information enables Citizen Participation.
13. $A, C \rightarrow D$: Accountability and Citizen Participation strengthen Democracy.
14. $D \rightarrow SW$: Democracy contributes to enhanced Social Wellbeing.
15. $T \rightarrow SW$: Transparency also leads directly to enhanced Social Wellbeing.
16. $R \rightarrow (D, T)$: Reforms enhance both Democracy and Transparency.

Let:

---

[60] See. Peralta, B. C. y Calvache, T. R. (2022). Una revisión histórica de la política social, Estado de bienestar y la emergencia de nuevos marcos discursivos en su construcción. Jurídicas, 19(1), 39-55. https://doi.org/10.17151/jurid.2022.19.1.3
[61] See. Coddou Mc Manus, A., & Smart Larraín, S. (2021). La transparencia y la no discriminación en el Estado de bienestar digital. Revista Chilena De Derecho Y Tecnología, 10(2), 301–332. https://doi.org/10.5354/0719-2584.2021.61034

Axioms Democracy, Transparency and Social Wellbeing = aDTSW

Then:

aDTSW = (D→T) ∧ (T→I) ∧ (I→A) ∧ (I→C) ∧ (A∧C→D) ∧ (D→SW) ∧ (T→SW) ∧ (R→(D∧T))

The set of axioms, denoted as aDTSW, defines a series of relationships among Democracy (D), Transparency (T), Information (I), Accountability (A), Citizen Participation (C) and Social Wellbeing (SW). These relationships can be summarized as follows: See Figure 6.

Figure 6
Exploration Axioms Democracy, Transparency and Social

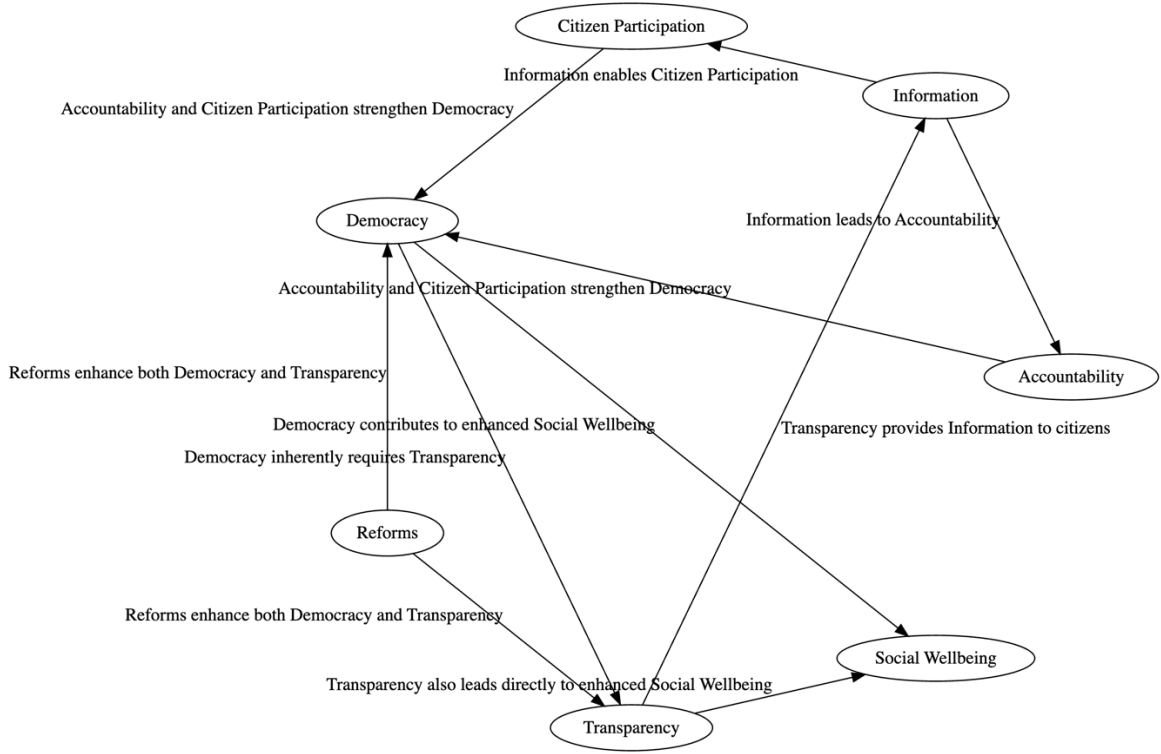

Source: Own elaboration

Democracy inherently requires Transparency because transparency provides information to citizens, which in turn leads to accountability and enables citizen participation. The combination of accountability and citizen participation strengthens democracy and democracy contributes to enhanced social wellbeing. Additionally, transparency directly leads to enhanced social wellbeing. Furthermore, implementing reforms enhances both democracy and transparency. In a concise notation, aDTSW captures these interdependencies.

**Inferential notation:**

$$D \xleftrightarrow{R} T \xrightarrow{I} (A, C) \longrightarrow D$$
$$D, T \longrightarrow S$$

**Explanation:**

1. Democracy and Transparency are mutually reinforcing, and Reforms ($R$) can enhance both ($D \xleftrightarrow{R} T$).
2. Transparency leads to the dissemination of Information ($T \xrightarrow{I} (A, C)$).
3. Information subsequently promotes both Accountability and Citizen Participation ($I \rightarrow (A, C)$).
4. Enhanced Accountability and Citizen Participation further bolster Democracy ($A, C \rightarrow D$).
5. Both Democracy and Transparency directly contribute to improved Social Wellbeing ($D, T \rightarrow S$).

In this model, the fundamental pillars, Democracy (D), Transparency (T) and Social Wellbeing (S), are intricately linked with additional factors such as Information (I), Accountability (A) and Citizen Participation (C). These relationships provide a holistic view, revealing how each component influences and is influenced by others, demonstrating their interdependence and the significance of each in reinforcing and enhancing the others.

## 2.1.1 Mathematical representation

**Variables:**

D: Level of Democracy
T: Level of Transparency
W: Social Welfare
C: Level of Trust and Credibility
E: Governance Effectiveness
other: Other relevant variables

**Parameters:**

α, β, γ, δ: Coefficients determining the relative influence of each variable on the others.

**Equations:**

Relationship between Democracy, Transparency and Social Wellbeing:

$$W = {_\alpha}D + {_\beta}T + {_\gamma}DT$$

This indicates that social welfare (W) is a function of the level of democracy (D), the level of transparency (T) and their interaction (DT).

Relationship between Transparency and Trust and Credibility:

$$C = {_\delta}T$$

This implies that trust and credibility (C) are proportional to the level of transparency (T).

Relationship between Transparency, Trust and Credibility and Governance Effectiveness:

$$E = {_\alpha}T + {_\beta}C + {_\gamma}TC$$

This shows that governance effectiveness (E) depends on transparency (T), trust and credibility (C) and their interaction (TC).

Initial Conditions and Bounds: It is essential to define initial values or limits for D, T, C and E to solve these equations. In addition, the coefficients $\alpha, \beta, \gamma$ and $\delta$ should be empirically determined through data collection and analysis.

This model seeks to quantify the relationships between democracy, transparency, social welfare, trust, credibility, and governance effectiveness.

## 2.1.2 Probabilistic model

Random Variables:

Let's consider the variables D, T, C and E as random variables, each with its own probability distribution.

- $D \sim p(D)$
- $T \sim p(T)$
- $C \sim p(C|D,T)$
- $E \sim p(E|D,T,C)$

**Equations with Probabilistic Elements:**

1. Relationship between Democracy, Transparency and Social Wellbeing:

$$W \sim p(W|D,T) = \alpha p(D) + \beta p(T) + \gamma p(D,T)$$

Where $p(W|D,T)$ is the joint probability of social welfare given D and T.

2. Relationship between Transparency and Trust and Credibility:

$$C \sim p(C|T) = \delta p(T)$$

Where $p(C|T)$ is the probability of the level of trust and credibility given T.

3. Relationship between Transparency, Trust and Credibility and Governance Effectiveness:

$$E \sim p(E|T,C) = \alpha p(T) + \beta p(C|T) + \gamma p(T,C|T)$$

Where $p(E|T,C)$ is the probability of governance effectiveness given T and C

**Expectations:**

Expectations of W, C and E can be calculated to obtain point estimates in the presence of uncertainty:

$$E[W] = \int W \cdot p(W|D,T)\, dW$$

$$E[C] = \int C \cdot p(C|T)\, dC$$

$$E[E] = \int E \cdot p(E|T,C)\, dE$$

The probabilistic model allows considering the inherent uncertainty in each of the variables and their interactions. Although it is more complex and requires more advanced estimation techniques (such as Bayesian inference), it provides a more realistic and robust framework for analyzing social and political systems.

## 2.2 Fundamental Postulates

**Postulate 1:** Democratic Essentiality

Democracy (D) is a necessary condition for the conjunction of Transparency (T) and Social Wellbeing (SW):
$$D \rightarrow (T \wedge SW)$$

**Postulate 2:** Inherent Information Flow

Transparency (T) implies the Inherent Flow of Information (I) to citizens:
$$T \rightarrow I$$

**Postulate 3:** Unavoidable Accountability

Inherent Information Flow (I) leads to Unavoidable Accountability (A) for systems and individuals:
$$I \rightarrow A$$

**Postulate 4:** Information-Enabled Engagement

The presence of Inherent Information Flow (I) and Unavoidable Accountability (A) ensures Information-Enabled Engagement (E) by citizens:
$$IA \rightarrow E$$

**Postulate 5:** Reinforced Democracy

The conjunction of Democracy (D) and Unavoidable Accountability (A) results in Reinforced Democracy (RD):
$$(D \wedge A) \rightarrow RD$$

**Postulate 6:** Wellbeing Enhancement

Democracy (D) leads to an increase in Social Wellbeing (SW) and Citizen Contentment (C):
$$D \rightarrow (SW \uparrow C)$$

**Postulate 7:** Transparent Wellbeing Correlation

Transparency (T) is associated with an increase in Social Wellbeing (SW):
$$T \rightarrow (SW \uparrow)$$

**Postulate 8:** Reformative Impacts

Implementing Reforms (R) results in an increase in Democracy (D) and Transparency (T):
$$R \rightarrow (D \uparrow \wedge T \uparrow)$$

**Postulate 9**: Wellbeing Interaction

The conjunction of Democracy (D) and Transparency (T) leads to a cumulative increase in Social Wellbeing (SW):
$$(D \wedge T) \rightarrow (SW \uparrow \uparrow)$$

**Postulate 10:** Transparent Trust

Transparency (T) is associated with an increase in Trust (Trust):

$$T \rightarrow (Trust \uparrow)$$

**Postulate 11:** Effective Governance

The conjunction of Transparency (T) and Trust (Trust) results in Effective Governance (EG):

$$(T \wedge Trust) \rightarrow EG$$

**Postulate 12:** Probabilistic Wellbeing Enhancement:

The conjunction of Democracy (D) and Transparency (T) probabilistically leads to an increase in Social Welfare (SW).

$$(D \wedge T) \rightarrow (SW \uparrow \sim P)$$

**Postulate 13:** Credibility Probability

Transparency (T) probabilistically leads to an increase in Trust (Trust):

$$T \rightarrow (Trust \uparrow \sim P)$$

**Postulate 14:** Effective Probability

The conjunction of Transparency (T) and Trust (Trust) probabilistically leads to Effective Governance (EG):
$$(T \wedge Trust) \rightarrow (EG \sim P)$$

Where:

| | |
|---|---|
| Citizen Contentment | (C) |
| Cumulative Increase in Social Wellbeing | (SW) |
| Democracy | (D) |
| Effective Governance | (EG) |
| Implementing Reforms | (R) |
| Information-Enabled Engagement | (E) |
| Inherent Flow of Information | (I) |
| Probabilistic Increase in Effective Governance | (EG) |
| Probabilistic Increase in Social Welfare | (SW) |
| Probabilistic Increase in Trust | (Trust) |
| Reinforced Democracy | (RD) |
| Social Wellbeing | (SW) |
| Transparency | (T) |
| Trust | (Trust) |
| Unavoidable Accountability | (A) |

We can represent these postulates together in a general notation as follows.

**Postulate 1**: Democracy is essential

The harmonious coexistence of Transparency (T) and Social Wellbeing (SW) requires democracy (D). This fundamental principle states that transparency and social wellbeing thrive in democratic societies when citizens actively participate in governance. Open and equitable processes in democracy lead to transparency in decision-making and policymaking. Citizens' active participation in molding society usually leads to policies and actions that improve population well-being.

**Postulate 2:** Inherent Information Flow

Transparency (T) promotes information (I) flow to citizens, according to this hypothesis. Citizens can access, understand and actively participate in government actions, policies and choices when information is transparent. The flow of information is crucial to informed citizenship and participatory governance.

**Postulate 3:** Unavoidable Accountability

It suggests that in a society characterized by Inherent Information Flow (I), evading accountability (A) becomes a challenging proposition. In essence, when information regarding actions and decisions is easily accessible, both individuals and institutions are more likely to be held answerable for their conduct. This postulate underscores the instrumental role of transparency in incentivizing accountability.

**Postulate 4:** Information-Enabled Engagement

A society with Inherent Information Flow (I) and Unavoidable Accountability (A) allows citizens to engage in Information-Enabled Engagement (E). Information empowers and motivates citizens to participate in democracy. Engagement can include voting, advocacy, or other civic activities.

**Postulate 5**: Reinforced Democracy

This postulate suggests that the fusion of Democracy (D) and Unavoidable Accountability (A) leads to the fortification of Reinforced Democracy (RD). In essence, when a democratic system is accompanied by a culture of accountability, it becomes more resilient and robust. Citizens trust that their voices matter and that those in positions of authority will be held responsible for their actions.

**Postulate 6:** Wellbeing Enhancement

Socioeconomic wellbeing and citizen satisfaction are linked to democracy (D). Citizens' engagement in decision-making typically leads to policies that benefit the public in democratic countries. This boosts the happiness of citizens.

**Postulate 7:** Transparent Wellbeing Correlation

This postulate asserts that Transparency (T) is correlated with an enhancement in Social Wellbeing (SW). In transparent societies, government actions and policies are visible, allowing citizens to hold leaders accountable for their impact on social wellbeing. Transparency can lead to policies that directly or indirectly elevate the quality of life and wellbeing of the population.

**Postulate 8:** Reformative Impacts

The implementation of Reforms (R) is anticipated to yield an upswing in both Democracy (D) and Transparency (T). Reforms encompass changes and enhancements made to existing systems and structures. When reforms are introduced, they often aim to augment democratic processes and render government actions more transparent.

**Postulate 9:** Wellbeing Interaction

It suggests that the simultaneous presence of Democracy (D) and Transparency (T) engenders a cumulative elevation in Social Wellbeing (SW). In democratic and transparent societies, there exists a synergistic effect on social wellbeing. The active involvement of citizens in governance and the visibility of government actions collectively contribute to an overarching improvement in social wellbeing.

**Postulate 10:** Transparent Trust

Transparency (T) is associated with a bolstering of Trust (Trust). When citizens have access to transparent information, their propensity to trust the government and its institutions increases. Trust in government is pivotal for the stability and effective operation of a democratic society.

**Postulate 11**: Effective Governance

The presence of both Transparency (T) and Trust (Trust) culminates in Effective Governance (EG). Effective governance materializes when government institutions are not only transparent but also command the trust of the public. This signifies that the government can efficiently and equitably execute its functions.

**Postulate 12:** Probabilistic Wellbeing Enhancement

This postulate suggests that the confluence of Democracy (D) and Transparency (T) introduces a probabilistic influence on the augmentation of Social Welfare (SW). It acknowledges that while democracy and transparency tend to foster social welfare, the outcome is probabilistic and can vary under different circumstances.

**Postulate 13:** Credibility Probability

Transparency (T) is projected to probabilistically instigate an upsurge in Trust (Trust). Transparency enhances the credibility of government actions and decisions, thus probabilistically nurturing trust among citizens.

**Postulate 14:** Effective Probability

The coexistence of Transparency (T) and Trust (Trust) probabilistically leads to Effective Governance (EG). This postulate acknowledges that the simultaneous presence of transparency and trust heightens the probability of effective governance, although it does not guarantee it.

Collectively, these postulates articulate the intricate interplay between democracy, transparency, accountability, trust and their far-reaching implications for various facets of society, including wellbeing, civic engagement and governance. See next Figure 7.

Figure 7
Interconnected Postulates: Democracy, Transparency and Social Wellbeing

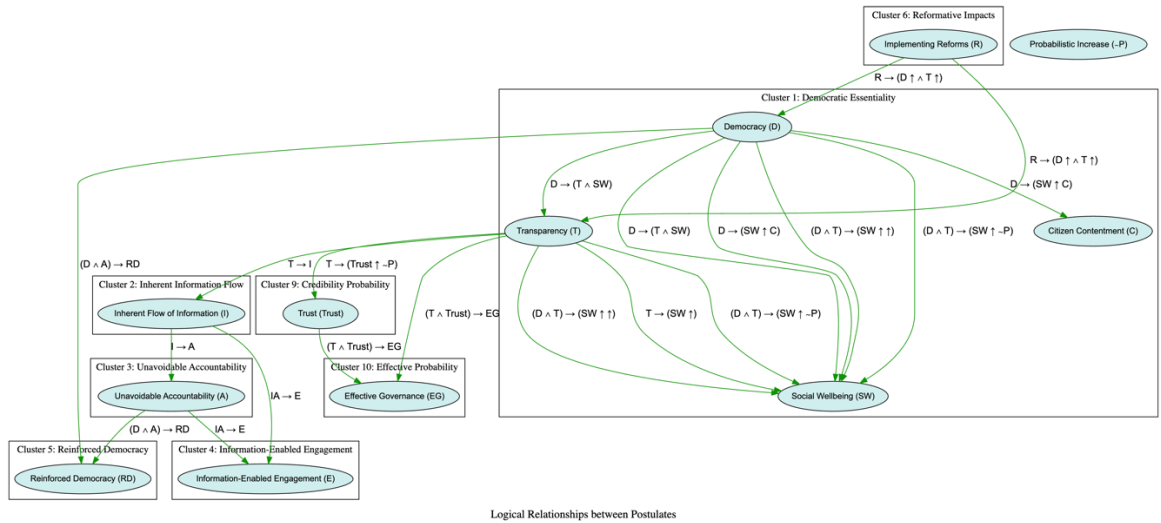

Source: Own elaboration

Let:
$$\text{Democracy, Transparency and Social Wellbeing} = pDTSW$$

Then:

$$pDTSW = (D \rightarrow (T \wedge SW)) \wedge (T \rightarrow I) \wedge (I \rightarrow A) \wedge ((I \wedge A) \rightarrow E) \wedge ((D \wedge A) \rightarrow RD) \wedge (D \rightarrow (SW \uparrow C)) \wedge (T \rightarrow (SW \uparrow)) \wedge (R \rightarrow (D \uparrow \wedge T \uparrow)) \wedge ((D \wedge T) \rightarrow (SW \uparrow \uparrow)) \wedge (T \rightarrow (Trust \uparrow)) \wedge ((T \wedge Trust) \rightarrow EG) \wedge ((D \wedge T) \rightarrow (SW \uparrow \sim P)) \wedge (T \rightarrow (Trust \uparrow \sim P)) \wedge ((T \wedge Trust) \rightarrow (EG \sim P))$$

The expression pDTSW represents a set of interconnected conditions that relate Democracy (D), Transparency (T), Social Wellbeing (SW), Information (I), Accountability (A), Citizen Participation (C), Reforms (R), Trust and Effective Governance (EG). These conditions establish that democracy requires transparency and contributes to social wellbeing, that transparency provides information and increases trust, and that information leads to accountability and citizen participation, strengthening democracy. Additionally, reforms enhance both democracy and transparency and the presence of transparency and trust results in effective governance with associated probabilities.

## Key Assumptions

These assumptions offer insights.

> **Assumption 1:** Democratic Stability

A stable democracy (D) is essential for the coexistence of both Transparency (T) and Social Wellbeing (SW). In other words, for transparency and social wellbeing to thrive, a democracy must be established and stable.

$$D \rightarrow (T \land SW)$$

> **Assumption 2:** Information Accessibility

Transparency (T) entails the accessibility of information (I) to citizens. When a government embraces transparency, it signifies that data concerning its actions and decisions is readily accessible to the public.

$$T \rightarrow I$$

> **Assumption 3:** Accountability Framework

When there is an inherent flow of information (I), it leads to an accountability framework (A). In essence, when information is accessible, individuals and institutions are more likely to be held accountable for their actions.

$$I \rightarrow A$$

> **Assumption 4:** Citizen Participation

The combination of inherent information flow (I) and accountability (A) encourages Information-Enabled Engagement (E) by citizens. When information is available and accountability is in place, citizens are more likely to engage actively in the democratic process.

$$(I \land A) \rightarrow E$$

> **Assumption 5:** Sustaining Democracy

The conjunction of Democracy (D) and Accountability (A) results in the reinforcement of Democracy (RD). In other words, when democracy is accompanied by a culture of accountability, it becomes more robust and resilient.

$$(D \land A) \rightarrow RD$$

> **Assumption 6:** Wellbeing Prioritization

Democracy (D) inherently leads to an improvement in Social Wellbeing (SW) and Citizen Contentment (C). In democratic societies, policies often prioritize the overall wellbeing of the population, leading to citizen satisfaction.

$$D \rightarrow (SW \uparrow C)$$

> **Assumption 7:** Wellbeing Impact

Transparency (T) is associated with an increase in Social Wellbeing (SW). When government actions and policies are transparent, citizens can hold leaders accountable for their impact on social wellbeing, leading to improvements.

$$T \rightarrow (SW \uparrow)$$

**Assumption 8:** Reform-driven Progress

Implementing Reforms (R) results in an increase in both Democracy (D) and Transparency (T). Reforms often aim to enhance democratic processes and make government actions more transparent.

$$R \rightarrow (D \uparrow \wedge T \uparrow)$$

**Assumption 9:** Enhanced Wellbeing

The combination of Democracy (D) and Transparency (T) leads to a cumulative increase in Social Wellbeing (SW). In democratic and transparent societies, there is a synergistic effect on social wellbeing.

$$(D \wedge T) \rightarrow (SW \uparrow \uparrow)$$

**Assumption 10:** Trust Building

Transparency (T) is correlated with an increase in Trust (Trust). When citizens have access to transparent information, they are more likely to trust the government and its institutions.

$$T \rightarrow (Trust \uparrow)$$

**Assumption 11:** Effective Governance Nexus

The conjunction of Transparency (T) and Trust (Trust) results in Effective Governance (EG). Effective governance occurs when government institutions are both transparent and trusted by the public.

$$(T \wedge Trust) \rightarrow EG$$

**Assumption 12:** Probabilistic Welfare

The combination of Democracy (D) and Transparency (T) probabilistically leads to an increase in Social Welfare (SW). It acknowledges that while democracy and transparency promote social welfare, the outcome is probabilistic and may vary in different circumstances.

$$(D \wedge T) \rightarrow (SW \uparrow \sim P)$$

**Assumption 13:** Credibility and Trust

Transparency (T) probabilistically leads to an increase in Trust (Trust) through the credibility of government actions, although not guaranteed in every situation.

$$T \rightarrow (Trust \uparrow \sim P)$$

**Assumption 14:** Probabilistic Effectiveness

The simultaneous presence of transparency and trust probabilistically increases the likelihood of Effective Governance (EG), but it doesn't guarantee it, recognizing that effectiveness can vary in different contexts.

$$(T \wedge Trust) \rightarrow (EG \sim P)$$

These explanations provide a comprehensive understanding of how each assumption relates to the dynamics of democracy, transparency, accountability, wellbeing, trust and governance.

We can represent this assumption together in a general notation as follows. See next Figure 8.

Figure 8
The Interconnected Assumptions: Democracy, Transparency and Social Wellbeing

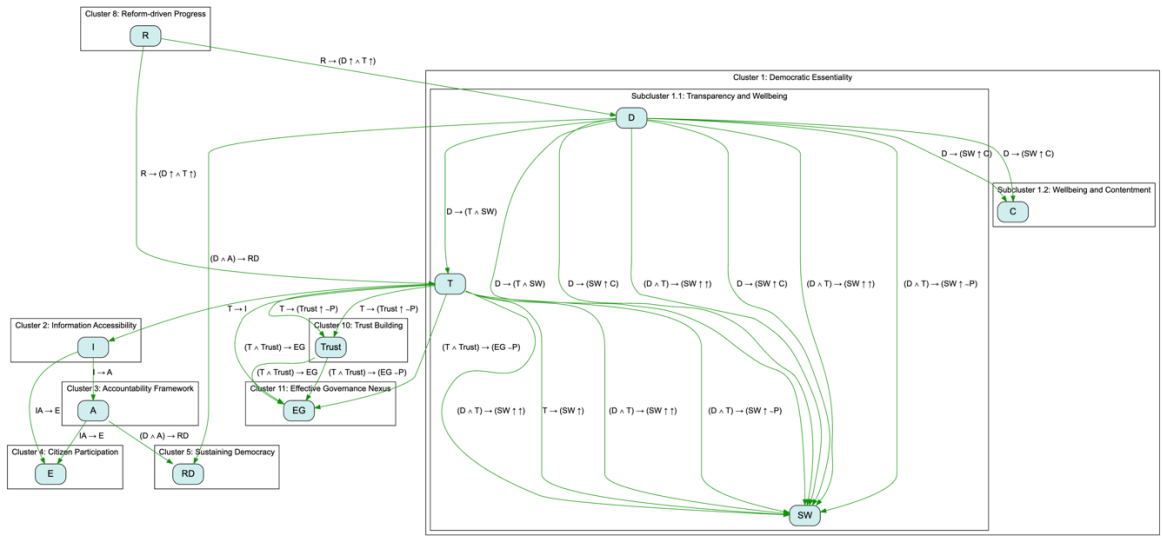

Source: Own elaboration

These assumptions underscore the evolving relationship between technology and governance in today's digital age. The tapestry of change, challenges, and opportunities they paint serves as a compass for the future trajectories of nations globally.

Let:

Assumptions on Democratic Stability, Governance an Social Wellbeing = aDSGSW

Then:

$$aDSGSW = GM \leftrightarrow (TA \lor TG \lor IC \lor JE \lor PSM \lor N \lor DA \lor FMI \lor S \lor CE \lor SE \lor CC \lor EG \lor GM \lor CL \lor AI \lor AR \lor AC)$$

Where:

| | |
|---|---|
| Democracy | (D) |
| Transparency | (T) |
| Social Wellbeing | (SW) |
| Information Accessibility | (I) |
| Accountability | (A) |
| Information-Enabled Engagement | (E) |
| Reforms | (R) |
| Trust | (Trust) |
| Effective Governance | (EG) |
| Probabilistic Welfare | (SW ↑ ~P) |
| Credibility and Trust | (Trust ↑ ~P) |
| Probabilistic Effectiveness | (EG ~P) |
| Probabilistic | (P) |

The equation aDSGSW represents a set of assumptions about the relationships between Democracy (D), Transparency (T), Social Wellbeing (SW) another factors. It suggests that these factors are interconnected, and the assumptions are based on various conditions (TA, TG, etc.). Additionally, there are probabilistic elements (↑ ~P) indicating that the outcomes may not always be certain and they depend on these conditions. So, it's like a way of expressing how different aspects of governance and wellbeing are linked in a complex system.

## 2.4 Foundational theorems: Exploring Democracy, Transparency and Social Wellbeing

**Theorem 1:** Democratic Stability Theorem

In a stable democracy (D), the coexistence of Transparency (T) and Social Wellbeing (SW) is guaranteed:
$$D \rightarrow (T \wedge SW)$$

**Theorem 2:** Information Accessibility Theorem

Transparency (T) implies the accessibility of information (I) to citizens:

$$T \rightarrow I$$

**Theorem 3:** Accountability Framework Theorem

The presence of an inherent flow of information (I) leads to the establishment of an accountability framework (A):

$$I \rightarrow A$$

**Theorem 4:** Citizen Participation Theorem

When there is both inherent information flow (I) and accountability (A), it ensures Information-Enabled Engagement (E) by citizens:

$$(I \wedge A) \rightarrow E$$

**Theorem 5:** Reinforced Democracy Theorem

The conjunction of Democracy (D) and Accountability (A) results in the reinforcement of Democracy (RD):
$$(D \wedge A) \rightarrow RD$$

**Theorem 6:** Wellbeing Prioritization Theorem

Democracy (D) inherently leads to an increase in Social Wellbeing (SW) and Citizen Contentment (C):
$$D \rightarrow (SW \uparrow C)$$

**Theorem 7**: Wellbeing Impact Theorem

Transparency (T) is associated with an increase in Social Wellbeing (SW):

$$T \rightarrow (SW \uparrow)$$

**Theorem 8**: Reform-driven Progress Theorem

Implementing Reforms (R) leads to an increase in both Democracy (D) and Transparency (T):
$$R \rightarrow (D \uparrow \wedge T \uparrow)$$

**Theorem 9:** Enhanced Wellbeing Theorem

The conjunction of Democracy (D) and Transparency (T) leads to a cumulative increase in Social Wellbeing (SW):

$$(D \land T) \rightarrow (SW \uparrow \uparrow)$$

**Theorem 10:** Trust Building Theorem

Transparency (T) is correlated with an increase in Trust (Trust):

$$T \rightarrow (Trust \uparrow)$$

**Theorem 11:** Effective Governance Nexus Theorem

The conjunction of Transparency (T) and Trust (Trust) results in Effective Governance (EG):

$$(T \land Trust) \rightarrow EG$$

**Theorem 12:** Probabilistic Welfare Theorem

The combination of Democracy (D) and Transparency (T) probabilistically leads to an increase in Social Welfare (SW):

$$(D \land T) \rightarrow (SW \uparrow \sim P)$$

**Theorem 13:** Credibility and Trust Theorem

Transparency (T) probabilistically leads to an increase in Trust (Trust) through the credibility of government actions:

$$T \rightarrow (Trust \uparrow \sim P)$$

**Theorem 14:** Probabilistic Effectiveness Theorem

The simultaneous presence of transparency and trust probabilistically increases the likelihood of Effective Governance (EG), recognizing that effectiveness can vary in different contexts:

$$(T \land Trust) \rightarrow (EG \sim P)$$

These theorems encapsulate the fundamental relationships and implications of democracy, transparency, accountability, wellbeing, trust and governance as described in the given axioms, postulates and assumptions.

We can represent these theorems together in a general notation as follows. See next Figure 9.

Figure 9
Theoretical Framework: Ten Key Theorems Shaping Transparent, Governance and Social Wellbeing

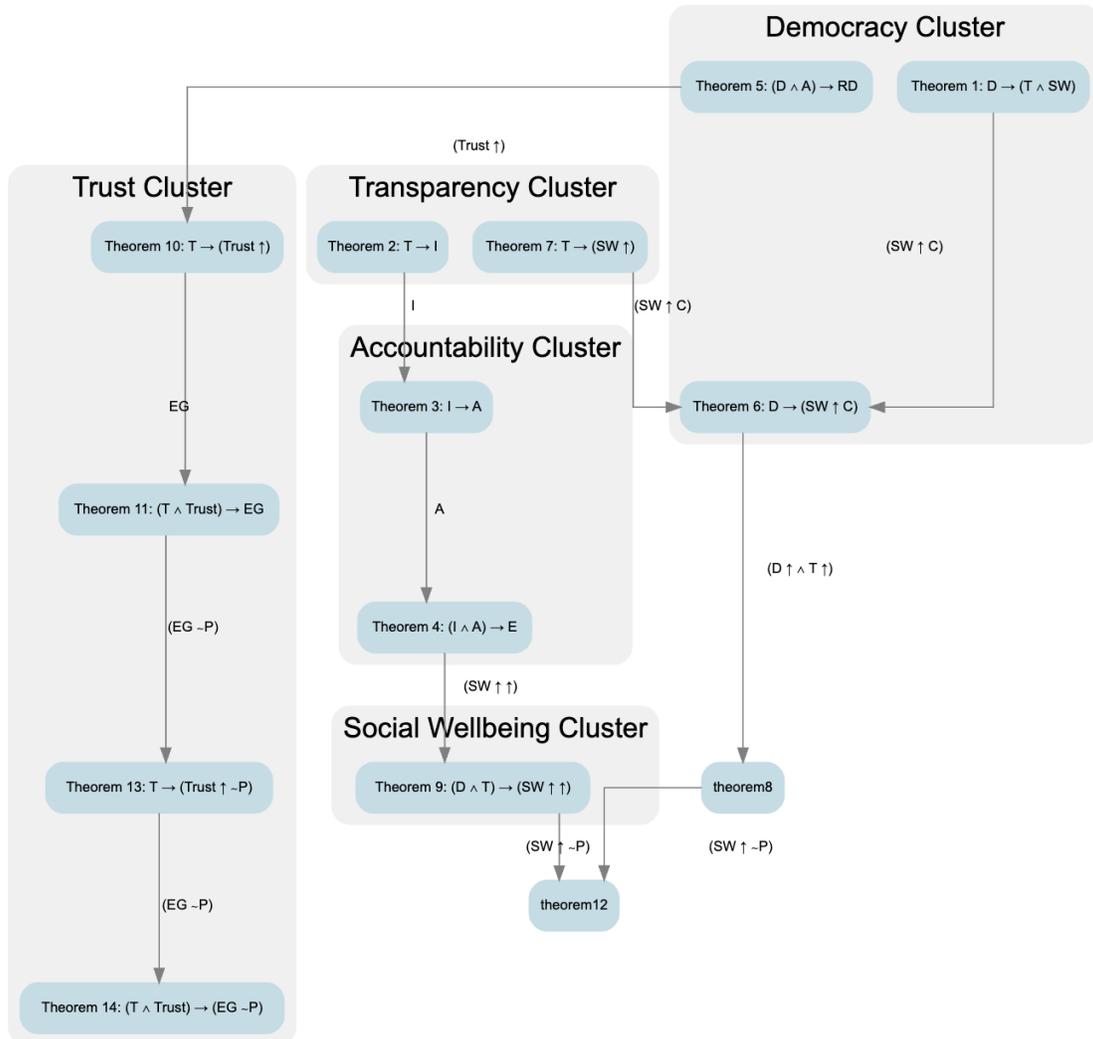

Source: Own elaboration

According to Figure 5, each of the expressed theorems is a logical statement that establishes a relationship between different concepts related to governance, technology, collaboration and transparency in the context of countries or nations. Here is the explanation of each one:

**Theorem 1:** Democratic Stability Theorem

In a stable democracy (D), this theorem asserts that there is a guaranteed coexistence of Transparency (T) and Social Wellbeing (SW). In other words, when a democracy is stable, it is expected that both transparency in government operations and the overall wellbeing of society will exist together.

**Theorem 2:** Information Accessibility Theorem

This theorem establishes that Transparency (T) implies that information (I) is accessible to citizens. In a transparent system, information should be readily available to the public, enabling citizens to access and comprehend government activities effectively.

**Theorem 3:** Accountability Framework Theorem

The existence of an intrinsic information flow (I) paves the way for the development of an accountability structure (A). This implies that when information naturally circulates within a system, it typically gives rise to systems and procedures that enforce responsibility upon individuals or entities for their conduct.

**Theorem 4:** Citizen Participation Theorem

When there is both inherent information flow (I) and accountability (A), this theorem states that it ensures Information-Enabled Engagement (E) by citizens. In such a context, citizens are more likely to engage actively and responsibly in the governance process due to the availability of information and mechanisms for accountability.

**Theorem 5:** Reinforced Democracy Theorem

This theorem suggests that when Democracy (D) and Accountability (A) coexist, it reinforces democracy itself (RD). In other words, when accountability mechanisms are in place within a democratic system, it strengthens the democratic principles.

**Theorem 6:** Wellbeing Prioritization Theorem

Democracy (D) inherently leads to an increase in Social Wellbeing (SW) and Citizen Contentment (C). This theorem highlights that democratic systems tend to contribute positively to the wellbeing and contentment of citizens.

**Theorem 7:** Wellbeing Impact Theorem

The concept of transparency (T) correlates with an enhancement of Social Wellbeing (SW). Within systems characterized by a heightened degree of transparency, a favorable influence on the overall welfare of society is anticipated.

**Theorem 8:** Reform-driven Progress Theorem

The implementation of Reforms (R) leads to an increase in both Democracy (D) and Transparency (T). This theorem emphasizes that reform initiatives are likely to result in improvements in democratic practices and transparency.

**Theorem 9:** Enhanced Wellbeing Theorem

The synergy between Democracy (D) and Transparency (T) results in a collective enhancement of Social Wellbeing (SW). When democracy and transparency collaborate harmoniously, their combined impact on the overall welfare of society is greater than the sum of their individual contributions.

**Theorem 10:** Trust Building Theorem

Transparency (T) is correlated with an increase in Trust (Trust). This theorem suggests that transparent governance fosters trust among citizens in the actions and decisions of the government.

**Theorem 11:** Effective Governance Nexus Theorem

The conjunction of Transparency (T) and Trust (Trust) results in Effective Governance (EG). When transparency and trust are both present, they form a nexus that contributes to effective governance.

**Theorem 12:** Probabilistic Welfare Theorem

The combination of Democracy (D) and Transparency (T) probabilistically leads to an increase in Social Welfare (SW). This theorem acknowledges that the presence of democracy and transparency increases the likelihood of improved social welfare, although outcomes may vary in different situations.

**Theorem 13**: Credibility and Trust Theorem

Transparency (T) probabilistically leads to an increase in Trust (Trust) through the credibility of government actions. Transparent actions by the government build credibility, which in turn enhances trust among the public.

**Theorem 14:** Probabilistic Effectiveness Theorem

The simultaneous presence of transparency and trust probabilistically increases the likelihood of Effective Governance (EG). This theorem acknowledges that the coexistence of transparency and trust increases the potential for effective governance, yet the degree of effectiveness remains contingent on contextual factors.

Together, these theorems delve into the intricate connections among democracy, transparency, accountability, wellbeing, trust, and governance, shedding light on how these components interplay and influence one another within the realm of politics and society. Consolidate these theorems into a single notation:

| | |
|---|---|
| Democracy | (D) |
| Transparency | (T) |
| Social Wellbeing | (SW) |
| Information Accessibility | (I) |
| Accountability Framework | (A) |
| Information-Enabled Engagement | (E) |
| Reinforcement of Democracy | (RD) |
| Citizen Contentment | (C) |
| Reforms | (R) |
| Trust | (Trust) |
| Effective Governance | (EG) |
| Probabilistic Increase | (P) |

Using the given information, the notations for the theorems can be represented as:

| | |
|---|---|
| Theorem 1: | (D) |
| Theorem 2: | $(T \rightarrow I)$ |
| Theorem 3: | $(I \rightarrow A)$ |
| Theorem 4: | $((I \land A) \rightarrow E)$ |
| Theorem 5: | $((D \land A) \rightarrow RD)$ |
| Theorem 6: | $(D \rightarrow (SW \uparrow C))$ |
| Theorem 7: | $(T \rightarrow (SW \uparrow))$ |
| Theorem 8: | $(R \rightarrow (D \uparrow \land T \uparrow))$ |
| Theorem 9: | $((D \land T) \rightarrow (SW \uparrow \uparrow))$ |
| Theorem 10: | $(T \rightarrow (Trust \uparrow))$ |
| Theorem 11: | $((T \land Trust) \rightarrow EG)$ |
| Theorem 12: | $((D \land T) \rightarrow (SW \uparrow \sim P))$ |

Let:

Theorems Shaping Democracy, Transparency, and Social Wellbeing = (tDTSW)

Then:

tDTSW = (D ∧ (T→I) ∧ (I→A) ∧ ((I∧A)→E) ∧ ((D∧A) → RD) ∧ (D→(SW↑C)) ∧ (T→(SW↑)) ∧ (R→(D↑∧T↑)) ∧ ((D∧T)→(SW↑↑)) ∧ (T→(Trust↑)) ∧ ((T∧Trust)→EG) ∧ ((D∧T)→(SW↑¬P)))

## 2.5 Theorems shaping Democracy, Transparency and Social Wellbeing (tDTSW)

Represent a comprehensive framework that establishes the intricate relationships between democracy, transparency, accountability, social wellbeing, trust and effective governance within the context of nations. This framework not only provides a theoretical foundation but also offers practical insights into how these elements interact and influence each other in the realm of politics and society.

At its core, tDTSW emphasizes the pivotal role of democracy (D) as a catalyst for positive change. It acknowledges that a stable democracy (D) is associated with both transparency (T) and social wellbeing (SW). This recognition underscores the importance of democratic governance in promoting the welfare of citizens and ensuring that government actions are conducted openly and honestly.

Transparency (T) emerges as a central theme in tDTSW, as it is intricately linked to various other components. The framework posits that transparency implies information accessibility (I), which, in turn, lays the foundation for an accountability framework (A). This sequence of events highlights the logical progression from open access to information to the establishment of mechanisms that hold individuals and entities accountable for their actions.

The tDTSW framework also underscores the reinforcing nature of democracy (D) and accountability (A). When accountability mechanisms are in place within a democratic system, they not only hold individuals accountable but also strengthen the democratic principles themselves (RD).

Additionally, democracy (D) is shown to have a positive impact on social wellbeing (SW) and citizen contentment (C). This observation aligns with the idea that democratic systems tend to prioritize the welfare of citizens and promote their overall satisfaction.

The relationship between transparency (T) and social wellbeing (SW) is explored in tDTSW, revealing that greater transparency correlates with enhanced social wellbeing. In societies characterized by a high degree of transparency, there is an expected improvement in the overall welfare of citizens.

Reforms (R) emerge as a powerful driver of positive change within the framework. The implementation of reforms is shown to lead to increased democracy (D) and transparency (T), underscoring the role of reform initiatives in enhancing governance practices and transparency.

Trust (Trust) is identified as a crucial element that is positively influenced by transparency (T). The framework suggests that transparent governance fosters trust among citizens in the government's actions and decisions, emphasizing the importance of trust-building in governance.

Moreover, the conjunction of transparency (T) and trust (Trust) is depicted as a key driver of effective governance (EG). When both transparency and trust are present, they form a nexus that contributes significantly to the effectiveness of governance, leading to better outcomes for society.

The probabilistic nature of some relationships in tDTSW acknowledges that while the presence of democracy (D) and transparency (T) increases the likelihood of improved social welfare (SW), outcomes may vary in different situations. It recognizes that external factors and contextual nuances can influence the degree of impact. The Theorems Shaping Democracy, Transparency and Social Wellbeing (tDTSW) framework provides a structured and interconnected view of the relationships between democracy, transparency, accountability, social wellbeing, trust and effective governance. See Figure 10.

Figure 10
Foundations, Postulates, Assumptions and Theorems: Democracy, Transparency and Social Wellbeing

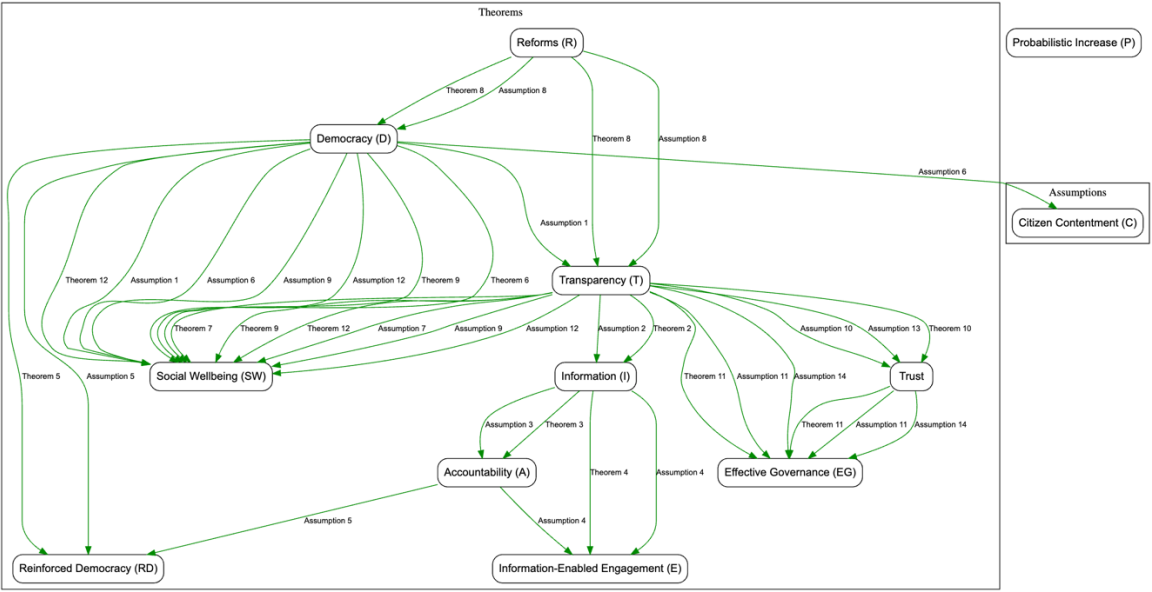

Source: Own elaboration

The Figure 9 illustrates the intricate web of relationships and assumptions that underlie the concepts of democracy, transparency and social wellbeing, along with their interdependencies. The theorems and assumptions outlined in this digraph help us understand the complex dynamics between these key elements.

Democracy (D), Transparency (T) and Social Wellbeing (SW) are central nodes, each with its own set of theorems and assumptions. The theorems highlight essential relationships, such as the influence of transparency on accountability (Theorem 3) or the reinforcement of democracy through information (Theorem 5).

Assumptions play a crucial role in connecting these nodes, shaping our understanding of their interplay. For instance, Assumption 1 posits that democracy influences both transparency and social wellbeing. Assumption 4 suggests that information and accountability together enable information-enabled engagement.

The digraph also illustrates how trust (Trust) is influenced by transparency and, in turn, contributes to effective governance (EG). These relationships underscore the critical role transparency plays in building trust and fostering effective governance.

In summary, this digraph provides a visual representation of the intricate network of theorems and assumptions that define the relationships between democracy, transparency, social wellbeing, trust and effective governance. It serves as a valuable tool for understanding the complex dynamics at play in the realm of governance and social wellbeing.

## 2.6 Preliminary conclusions

In Chapter 2, the tDTSW framework offers a comprehensive perspective on the intricate dynamics between democracy, transparency, and social wellbeing within the context of politics and society. It accentuates the central role of democracy as a catalyst for constructive transformation and highlights the critical significance of transparency, accountability, trust, and effective governance in fostering the overall welfare of citizens.

Beyond its theoretical value, this framework extends its practical implications to guide policymakers and stakeholders. It underscores the imperative of transparent governance, the imperative for reform implementation and the necessity of nurturing trust to enhance the efficacy of democratic systems and elevate the societal welfare.

While recognizing the inherent complexity and contextual nuances that may influence certain relationships, this framework stands as a valuable tool for illuminating the intricate tapestry of interconnections among democracy, transparency, and social wellbeing. By contemplating these theorems and assumptions, policymakers are equipped to make well-informed decisions that reinforce democratic principles and advance the prosperity of society.

# Chapter 3
# Fuzzy logic and Its role in Model tDTSW:
# An introduction to handling imprecise information

Fuzzy logic, a powerful tool for handling imprecise and uncertain information, plays a pivotal role in our exploration of the complex relationship between Democracy (D), Transparency (T) and Social Wellbeing (SW). In this introductory summary, we will delve into the key components of our model and its implications for understanding and evaluating these critical variables.

At the foundation of our model lies the concept of fuzzy sets. These sets are employed to represent Democracy (D), Transparency (T) and Social Wellbeing (SW), allowing us to move away from strict numerical categorizations and embrace the inherent vagueness and fuzziness in these real-world phenomena. Each variable is associated with fuzzy labels[62] such as "Low," "Medium," and "High," providing a qualitative framework that aligns more naturally with the subtleties of these concepts.

To bridge the gap between qualitative labels and numerical values, we introduce membership functions.[63] These mathematical functions assign degrees of membership to each label within the fuzzy sets, offering a systematic way to quantify how well a particular label corresponds to a numerical value. For instance, the membership function µ(Low_D) elucidates the degree to which a value belongs to the "Low" category of Democracy (D).

The heart of our model is constructed through the formulation of fuzzy rules. These rules articulate the intricate interactions between Democracy (D), Transparency (T) and Social Wellbeing (SW) across nine distinct scenarios.[64] Within each rule, we employ fuzzy operators such as "and" (AND) to combine the membership functions of Democracy (D) and Transparency (T), ultimately determining the degree of membership of Social Wellbeing (SW).

---

[62] Fuzzy sets are defined for each variable with labels that include "Low," "Medium," and "High." These sets represent qualitative levels of the variables.

[63] Membership functions are used to assign degrees of membership to each label in the fuzzy sets. These functions describe how the labels relate to numerical values.

[64] Expected Probability Scenarios, each scenario is represented as a combination of fuzzy values for D and T in the corresponding fuzzy rules. These scenarios are different combinations of ratings in D and T that can occur.

## 3.1 General model Democracy, Transparency and Social Wellbeing (tDTSW)

The general model uses fuzzy logic to evaluate how different combinations of ratings in D and T influence Social Wellbeing (SW). It allows for calculating the fuzzy value of SW for any specific set of fuzzy ratings for D and T. Variables are the fundamental elements in the model and represent key aspects of the analysis. D refers to Dimension, T to Subdimension and SW to Social Wellbeing, which is the desired outcome.

Fuzzy sets assign labels to the numerical values of the variables to express qualitative levels such as "Low," "Medium," and "High."

Membership functions are curves that relate the labels to numerical values and describe how degrees of membership are assigned to those labels. For example, a membership function for "Low" in D is expressed as μ(Low_D). Fuzzy rules represent the relationships between the variables D, T and SW in the 9 possible scenarios. Each rule is expressed with fuzzy operators like "and" (AND) to combine the degrees of membership of the labels of D and T and calculate the degree of membership of SW.

Expected probability scenarios are specific combinations of ratings in D and T that are evaluated using the fuzzy rules to determine the fuzzy value of SW.

The general model integrates all these elements and uses fuzzy logic to assess the potential impact of different combinations of dimensions and subdimensions on Social Wellbeing (SW), enabling more informed decisions based on data uncertainty and vagueness.

### 3.1.1 Fuzzy Rules model Democracy, Transparency and Social Wellbeing (tDTSW)

Fuzzy rules are defined to represent how the variables D and T affect the variable SW in each of the 9 possible scenarios. We use fuzzy operators like "and" (AND) to combine the membership functions of D and T in each rule.

### 3.1.2 Rules for scenarios of Social Wellbeing

Rule 1: If D is Low and T is Low, then SW is Low with a degree of membership P(SW = Low | D, T).

Rule 2: If D is Low and T is Medium, then SW is Medium with a degree of membership P(SW = Medium | D, T).

Rule 3: If D is Low and T is High, then SW is High with a degree of membership P(SW = High | D, T).

Rule 4: If D is Medium and T is Low, then SW is Medium with a degree of membership P(SW = Medium | D, T).

Rule 5: If D is Medium and T is Medium, then SW is Medium with a degree of membership P(SW = Medium | D, T).

Rule 6: If D is Medium and T is High, then SW is High with a degree of membership P(SW = High | D, T).

Rule 7: If D is High and T is Low, then SW is High with a degree of membership P(SW = High | D, T).

Rule 8: If D is High and T is Medium, then SW is High with a degree of membership P(SW = High | D, T).

Rule 9: If D is High and T is High, then SW is High with a degree of membership P(SW = High | D, T).

## 3.2 Fuzzy logic and Social Wellbeing assessment based on Democracy and Transparency

Fuzzy logic, an advanced mathematical tool designed to handle ambiguity and imprecise information, has garnered extensive utility across diverse domains such as control systems, artificial intelligence and the optimization of decision-making processes. Notably, fuzzy logic offers a valuable approach for evaluating social wellbeing (SW) within intricate systems characterized by the interplay of multiple variables that collectively shape the holistic state of societal well-being.

In this context, two critical variables often considered are democracy (D) and transparency (T). These variables are crucial in understanding and evaluating the functioning of governments and organizations. This article explores a set of fuzzy inference rules that link democracy, transparency and social welfare, providing a systematic approach for assessing SW in different scenarios.

### 3.2.1 Fuzzy logic and membership functions in model Democracy, Transparency and Social Wellbeing (tDTSW)

Before delving into the specific fuzzy inference rules, it's essential to understand the basic concepts of fuzzy logic, including membership functions. Fuzzy logic allows us to represent and manipulate uncertain or imprecise information. In the context of SW assessment, we use fuzzy sets to represent linguistic values such as "Low," "Medium," and "High." Each of these linguistic values is associated with a membership function that defines the degree to which an input (e.g., democracy or transparency) belongs to that category.

In our analysis, we use the following linguistic values for SW:

Low (SW = Low)
Medium (SW = Medium)
High (SW = High)

Similarly, we apply the same linguistic values to democracy (D) and transparency (T). Membership functions for each linguistic value in D, T and SW are defined based on domain knowledge and expert input. These membership functions determine the degree of membership (P) for each linguistic value in the corresponding set.

### 3.2.2 Fuzzy inference rules for Model (tDTSW)

Now, let's examine the nine fuzzy inference rules that connect D, T and SW. Each rule defines how SW is influenced by the values of democracy and transparency. We'll discuss each rule individually.

**Rule 1:** If D is Low and T is Low, then SW is Low with membership degree P(SW = Low | D, T).

This rule addresses a scenario where both democracy and transparency are low. In such a case, the membership degree of SW being low is determined by the degree of membership of both D and T being low. The rule reflects the idea that when both democracy and transparency are minimal, social welfare is likely to be low. See next Figure 11.

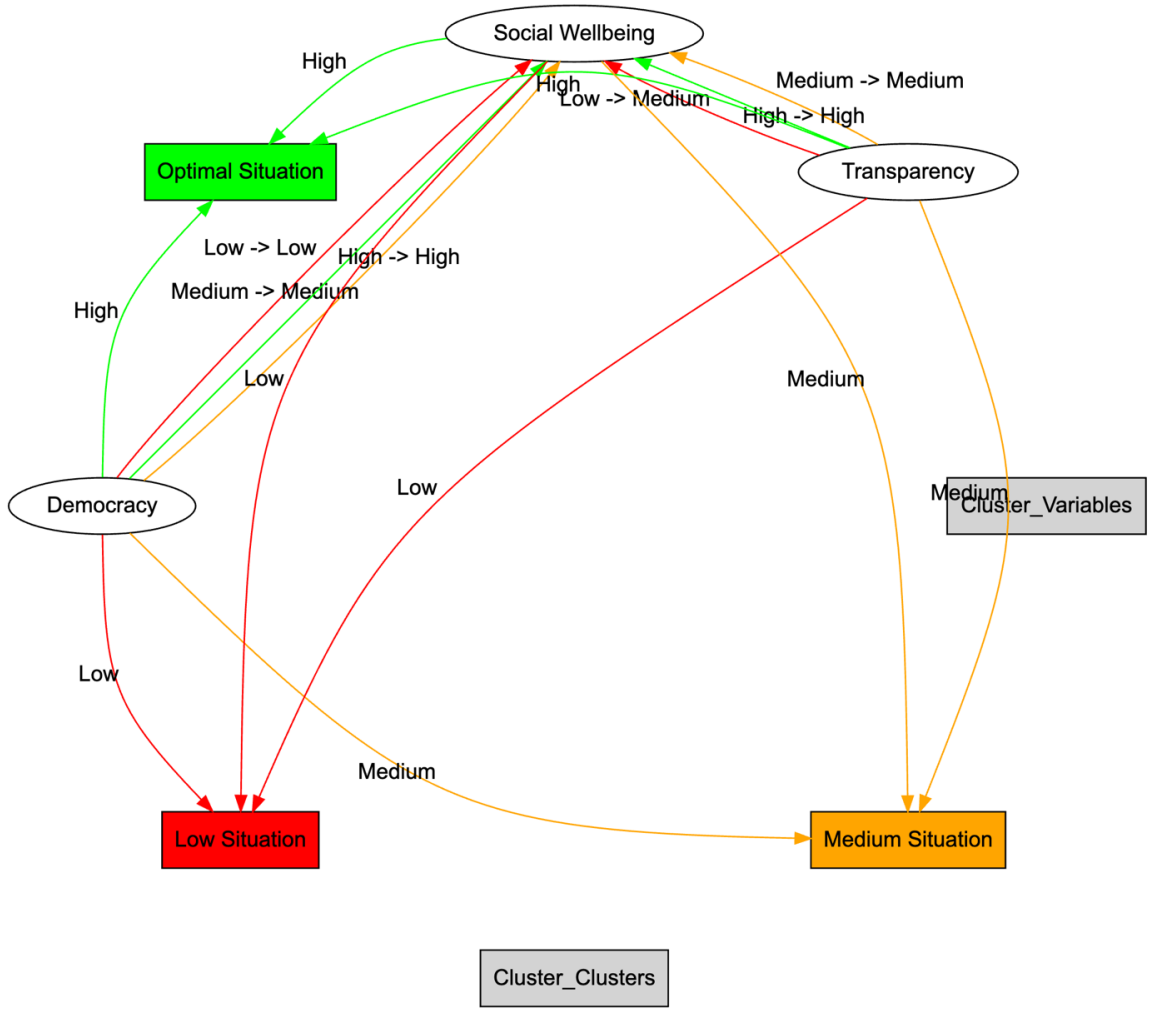

Figure 11
Fuzzy inference rules within the Model (tDTSW)

Source: Own elaboration

**Rule 2:** If D is Low and T is Medium, then SW is Medium with membership degree P(SW = Medium | D, T).

In this rule, we consider a situation where democracy is low, but transparency is at a medium level. The membership degree of SW being medium is influenced by the membership degrees of D being low and T being medium. This reflects the idea that moderate transparency can partially compensate for low democracy in terms of social welfare.

**Rule 3:** If D is Low and T is High, then SW is High with membership degree P(SW = High | D, T).

When democracy is low, but transparency is high, this rule suggests that social welfare is likely to be high. The membership degree of SW being high is determined by the membership degrees of D being low and T being high. High transparency can offset the negative effects of low democracy on social welfare.

**Rule 4: If** D is Medium and T is Low, then SW is Medium with membership degree P(SW = Medium | D, T).

This rule considers a scenario where democracy is at a medium level, but transparency is low. The membership degree of SW being medium depends on the membership degrees of D being medium and T being low. It implies that when democracy is moderate but transparency is lacking, social welfare tends to be moderate as well.

**Rule 5:** If D is Medium and T is Medium, then SW is Medium with membership degree P(SW = Medium | D, T).

In this rule, both democracy and transparency are at moderate levels, leading to the conclusion that social welfare is also moderate. The membership degree of SW being medium is determined by the membership degrees of D being medium and T being medium.

**Rule 6: If** D is Medium and T is High, then SW is High with membership degree P(SW = High | D, T).

When democracy is at a medium level and transparency is high, this rule suggests that social welfare is likely to be high. The membership degree of SW being high is influenced by the membership degrees of D being medium and T being high. High transparency can have a positive impact on social welfare in situations where democracy is moderately maintained.

**Rule 7:** If D is High and T is Low, then SW is High with membership degree P(SW = High | D, T).

This rule addresses a scenario where democracy is high, but transparency is low. The membership degree of SW being high is determined by the membership degrees of D being high and T being low. It suggests that strong democratic institutions can compensate for low transparency in terms of social welfare.

**Rule 8:** If D is High and T is Medium, then SW is High with membership degree P(SW = High | D, T).

In this rule, both democracy and transparency are at moderate levels, leading to the conclusion that social welfare is high. The membership degree of SW being high is influenced by the membership degrees of D being high and T being medium. This rule reflects the idea that a combination of strong democracy and moderate transparency leads to high social welfare.

**Rule 9:** If D is High and T is High, then SW is High with membership degree P(SW = High | D, T).

When both democracy and transparency are high, this rule suggests that social welfare is likely to be high. The membership degree of SW being high depends on the membership degrees of D being high and T being high. It reflects the idea that strong democracy and high transparency contribute positively to social welfare.

## Preliminary conclusions

This Chapter 3 has provided an in-depth exploration of the role of fuzzy logic in our model for assessing the complex relationship between Democracy (D), Transparency (T) and Social Wellbeing (SW). Fuzzy logic offers a powerful framework for handling imprecise and uncertain information, allowing us to move beyond rigid numerical categorizations and embrace the inherent vagueness in these real-world phenomena.

We have introduced the fundamental components of our model, starting with fuzzy sets that represent D, T and SW using qualitative labels such as "Low," "Medium," and "High." Membership functions have been employed to systematically quantify the degree of membership of each label, bridging the gap between qualitative and numerical values. Fuzzy rules, articulated through operators like "and" (AND), provide a framework for understanding how D and T interact to influence SW across nine distinct scenarios.

Our general model, tDTSW, integrates these elements to assess the impact of different combinations of dimensions and subdimensions on Social Wellbeing. By employing fuzzy logic, we can make more informed decisions while considering the inherent uncertainty and vagueness present in real-world data.

Additionally, we have discussed the nine fuzzy inference rules that define the relationships between D, TandSW in various scenarios. These rules offer a systematic way to evaluate SW based on the levels of democracy and transparency, providing valuable insights into how these critical variables interact.

Fuzzy logic serves as a robust tool for handling imprecise information and our model tDTSW leverages its capabilities to assess the intricate dynamics of Democracy, Transparency and Social Wellbeing. Through the amalgamation of the conceptual groundwork established in this chapter with real-world empirical data and practical applications, our objective is to enhance our comprehension of how these variables collectively influence the overall welfare of societies and provide valuable insights for decision-makers. This chapter serves as the foundation upon which we will build in the following sections, delving more profoundly into the tangible applications and ramifications of our model.

# Chapter 4
# Fuzzy inference process in tDTSW model

In the preceding chapter, we laid the groundwork for our tDTSW model, which examines the intricate interplay between Democracy (D), Transparency (T) and Social Wellbeing (SW) using the power of fuzzy logic. We discussed the essential components of our model, from fuzzy sets to membership functions and fuzzy rules, all designed to handle the inherent vagueness and uncertainty in real-world data.

Now, in Chapter 4, we dive deeper into the heart of our model, exploring the fuzzy inference process that underpins tDTSW. This chapter delves into the systematic application of the nine fuzzy inference rules that define the relationships between D, TandSW in various scenarios.

We will unravel how democracy and transparency levels interact and combine to influence social wellbeing, providing valuable insights into the complex dynamics at play. Through the utilization of fuzzy logic, we can make informed decisions that consider the nuanced nature of these critical variables.

Join us as we embark on a journey through the intricate world of fuzzy inference in the tDTSW model, where we unlock the potential to enhance our comprehension of how these variables collectively shape the welfare of societies. This chapter serves as a pivotal step forward, laying the foundation for the practical applications and real-world implications that we will explore in the subsequent sections.

## 4.1 Analyzing the graph model (tDTSW)

The fuzzy inference process involves evaluating these nine rules to determine the final membership degree of each linguistic value of SW (Low, Medium, and High) based on specific values of D and T. To do this, we consider the following steps:

   1.- Input values for D and T are provided.

   2.- The membership degrees for D and T in their respective linguistic values (Low, Medium, High) are calculated based on their membership functions.

   3.- For each rule, the minimum of the membership degrees of D and T is determined, representing the degree to which the rule is satisfied.

Using the minimum degree obtained in step 3, the corresponding membership degree for SW in the rule's conclusion is determined. This process is repeated for all nine rules and scenarios. See Figure 12.

Figure 12
Fuzzy inference rules for SW (Social Wellbeing) based on D (Democracy)
and T (Transparency) in Model (tDTSW)

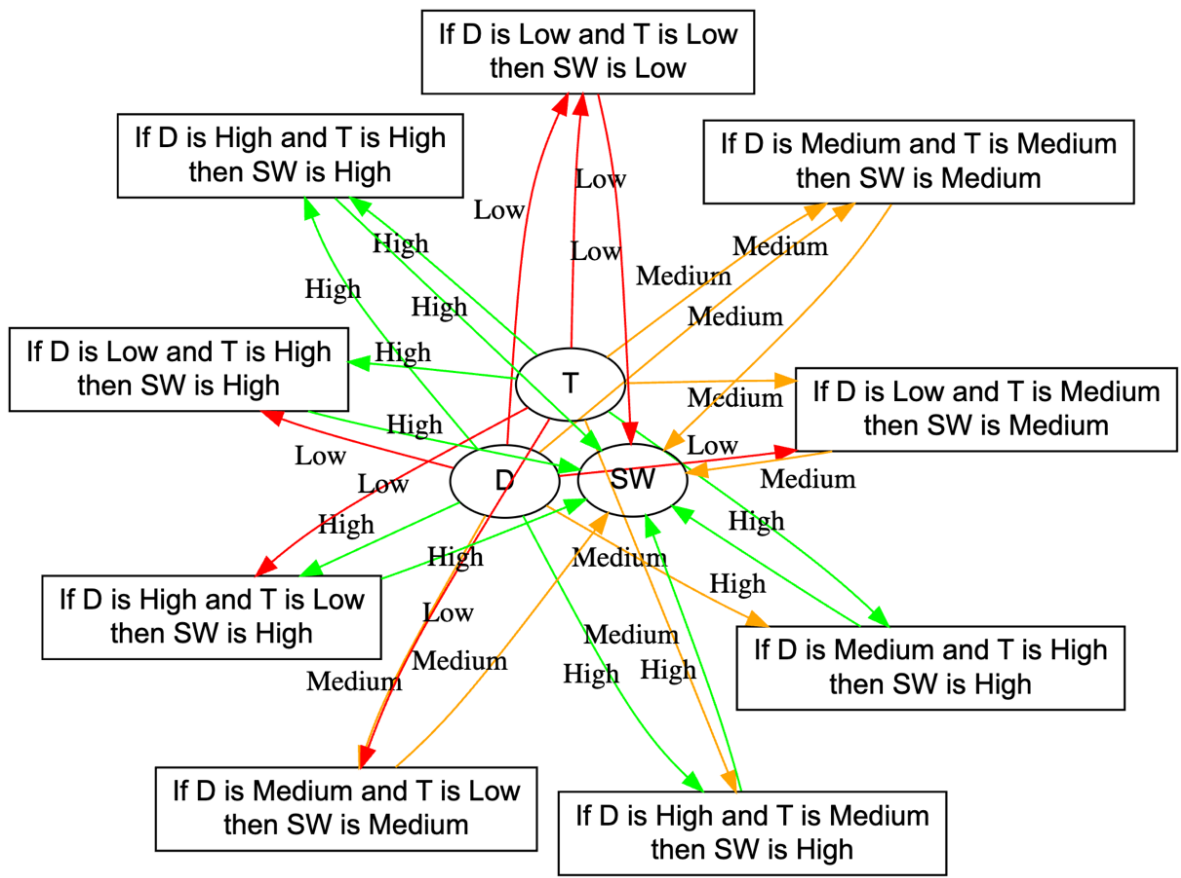

Source: Own elaboration

The final membership degrees for SW in its linguistic values (Low, Medium, High) are aggregated using the maximum operator to obtain the overall degree to which each linguistic value of SW applies. The final step provides us with a clear assessment of social welfare (SW) under the given conditions of democracy (D) and transparency (T). The degree of membership for each linguistic value of SW indicates the level of confidence or certainty in the assessment.

From the Figure 7, The line colors in this graphical representation play a crucial role in elucidating the relationships between variables and rules within the framework of fuzzy logic. Consider a fuzzy logic system with nine distinct scenarios indexed as R1, R2, ..., R9, each corresponding to a unique set of D and T levels. In this system, the color of the lines (R) conveys the influence ($\mu\_R(x)$) of D and T on SW.

1. Red (R): Denotes a low or weak influence, quantified as $\mu\_R(x)$ for a given rule R and input (x) in the universe of discourse. In this context, when a red line is observed, it implies that $\mu\_R(x)$ is minimal or negligible, signifying that the rule R has a low impact on the variable it pertains to.

2. Orange (O): Indicates a moderate or intermediate influence, represented as $\mu\_O(x)$ for rule R and input x. An orange line suggests that $\mu\_R(x)$ holds some degree of relevance, although it does not exert the strongest influence among the rules governing the system.

3. Green (G): Reflects a high or strong influence, defined as $\mu\_G(x)$ for rule R and input x. When a green line is encountered, it signifies that $\mu\_R(x)$ is substantial, emphasizing the significant impact of rule R on the variable it references.

This graphical representation elucidates the impact of different levels of democracy and transparency on social welfare using precise mathematical notation and fuzzy logic principles. The color-coded lines represent the varying degrees of influence, aiding in a rigorous scientific interpretation of the relationships within the system, which can inform decision-making and policy formulation.

## 4.2 Mathematically interpreting the scenarios' graph Model (tDTSW)

Suppose a fuzzy logic system has nine situations, R1, R2, … , R9, each with its own D and T levels. This system uses line color (R) to indicate the impact of D and T on SW ($\mu\_R(x)$).

**Scenario** (R1), if μ D and μ T are both "Low" (red), then μ SW is also "Low" (red). Low democracy (D) and transparency (T) levels result in low social welfare (SW) levels (μ SW is red), indicating a limited influence of these variables.

**Scenario** (R2), Low democracy (μ D in red) and medium transparency (μ T in orange) yield moderate social welfare (μ SW in orange), indicating a moderate influence.

**Scenario** (R3), low democracy (μ D in red) and high transparency (μ T in green) lead to high social welfare (μ SW in green), indicating a considerable influence.

**Scenario 4 (**R4), If μ D is "Medium" (orange) and μ T is "Low" (red), then μ SW is likewise "Medium" (orange). Moderate democracy (μ D orange) and low transparency (μ T red) lead to moderate social welfare (μ SW), showing moderate influence.

**Scenario 5** (R5), If $\mu\_D$ and μ T are "Medium" (orange), then $\mu\_{SW}$ is also "Medium" (orange). When democracy and transparency are moderate (μ D and μ T are orange), social welfare (μ SW) is moderate, indicating moderate impact.

**Scenario 6** (R6), If μ D is "Medium" (orange) and μ T is "High" (green), then μ SW is "High" (green). Moderate democracy (μ D orange) and high transparency (μ T green) indicate high social welfare (μ SW), indicating significant influence.

**Scenario 7** (R7): If μ D is "High" (green) and μ T is "Low" (red), μ SW is "High" (green). Higher democracy (μ D is green) and lower transparency (μ T is red) lead to higher social welfare (μ SW), demonstrating important influence.

**Scenario 8** (R8), If μ D is "High" (green) and μ T is "Medium" (orange), μ SW is "High" (green). High levels of democracy (μ D is green) and moderate transparency (μ T is orange) correlate with high social welfare (μ SW), indicating significant influence.

**Scenario 9** (R9), If μ D and μ_T are both "High" (green), then μ_SW is also "High" (green). High levels of democracy and transparency (μ_D and μ_T are green) are associated with high social welfare (μ_SW), indicating a significant impact.

Consider a fuzzy logic system with nine scenarios indexed as R1, R2, ..., R9, each characterized by specific membership degrees of democracy ($\mu D$) and transparency ($\mu T$). These scenarios influence the membership degree of social welfare ($\mu\_SW$). We can express this relationship as follows:

For each scenario $R_i$, where i ranges from 1 to 9:

$$\mu SW(R_i) = f(\mu D(R_i), \mu T(R_i))$$

Here,

| | |
|---|---|
| $\mu SW(R_i)$ | represents the membership degree of social welfare for scenario $R_i$ |
| $\mu D(R_i)$ | represents the membership degree of democracy for scenario $R_i$ |
| $\mu T(R_i)$ | represents the membership degree of transparency for scenario $R_i$ |
| $f(\mu D(R_i), \mu T(R_i))$ | is a function that determines the membership degree of social welfare based on the membership degrees of democracy and transparency for scenario $R_i$ |

The function $f(\mu D(R_i), \mu T(R_i))$ can be defined differently for each scenario $R_i$ to capture the specific relationships described in the scenarios.

## 4.3 Results of the model's (tDTSW) scenario graph

In this analysis, we explore three key variables: Democracy ($\mu_D$), Transparency ($\mu_T$) and Social Well-being ($\mu_{SW}$), each categorized as "Low," "Medium," or "High." These variables illustrate their influence across nine diverse scenarios in the Model (tDTSW). Through graphical representation, we examine how these factors shape the outcomes in different contexts. The colors "red," "orange," and "green" indicate the respective membership levels, providing valuable insights into the varying degrees of impact on each scenario. See Figure 13.

Figure 13
Influence of Social Wellbeing (SW) on Scenarios in the Model (tDTSW)

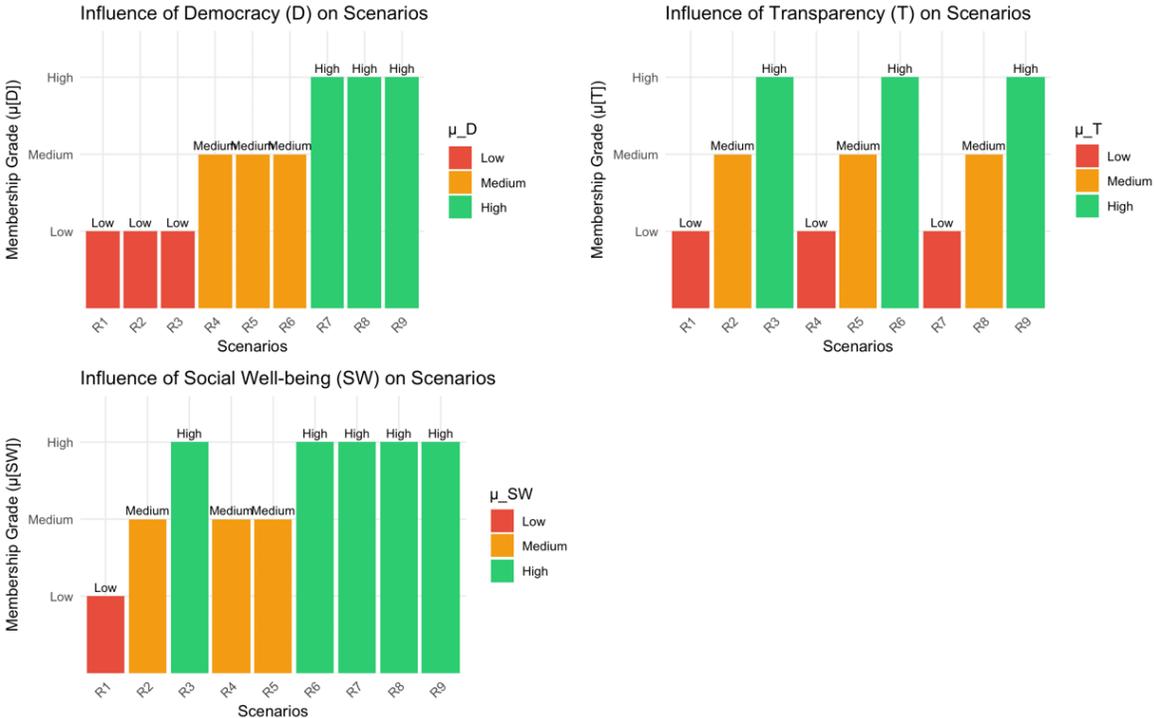

Source: Own elaboration

This exploration helps us understand the complex interplay between democracy, transparency, and social well-being within the model, aiding in informed decision-making and strategic planning.

According to Figure 12, it is evident that:

> **Variable: Democracy (µ D)**
>
> This variable shows the degree of "Democracy" membership in various circumstances. Membership might be "Low," "Medium," or "High," signifying democracy's impact.
>
>> Graphical results: Scenarios are on X, "Democracy" membership on Y. Colors represent membership levels ("Low" red, "Medium" orange, "High" green).

**Variable: (Transparency μ T)**

A variable explanation: The "Transparency" variable indicates "Transparency" membership in various contexts. Membership may be "Low," "Medium," or "High," depending on transparency, like the preceding variable.

Graphical results: The graph demonstrates how "Transparency" influences nine situations. The X-axis shows scenarios, and the Y-axis shows "Transparency" membership. Like before, colors denote membership levels ("Low" in red, "Medium" in orange and "High" in green). This graph shows how transparency changes between contexts.

**Variable: Social Wellbeing ( μ SW)**

A variable explanation: In several situations, the "Social Well-being" variable indicates membership in the idea. Membership may be "Low," "Medium," or "High," depending on social well-being, like the other characteristics.

Graphical results: The graph demonstrates how "Social Well-being" affects nine situations. The X-axis shows scenarios, and the Y-axis shows "Social Well-being" membership. As with prior graphs, colors denote membership levels ("Low" in red, "Medium" in orange and "High" in green). The graph shows how social well-being changes by circumstance.

These graphs and variables show how democracy, transparency and social wellbeing affect situations. Each graph shows variable variation by scenario and identifies patterns and trends that may be useful for decision-making and understanding how these factors affect results.

## 4.4 Unlocking the Potential of the tDTSW Model: Practical, Theoretical and Conceptual Insights

The mathematical interpretation of the scenarios' graph: Model (tDTSW) provides valuable insight into the influence of democracy, transparency, and social well-being in nine different situations. This analysis has multiple practical, theoretical, and conceptual utilities and it can also support well-being as a scenario for public policy in various ways. See next Figure 14.

### 4.4.1 Practical utility

*Policy decision-making:* The analysis of the model (tDTSW) allows policymakers to understand how policies related to democracy and transparency can impact social well-being. This can help design more effective policies and prioritize areas for improvement.

*Evaluation of existing policies*: Existing policies related to democracy and transparency can be evaluated using this model to determine their impact on social well-being. This is essential for adjusting and improving current policies.

*Identification of regional priorities:* Analyzing variations in democracy, transparency and social well-being across different scenarios can help identify regions or situations that require special attention.

*Strategic planning:* Governmental and non-governmental organizations can use this analysis to guide their strategic planning and allocate resources more effectively in areas that promote social well-being.

Figure 14

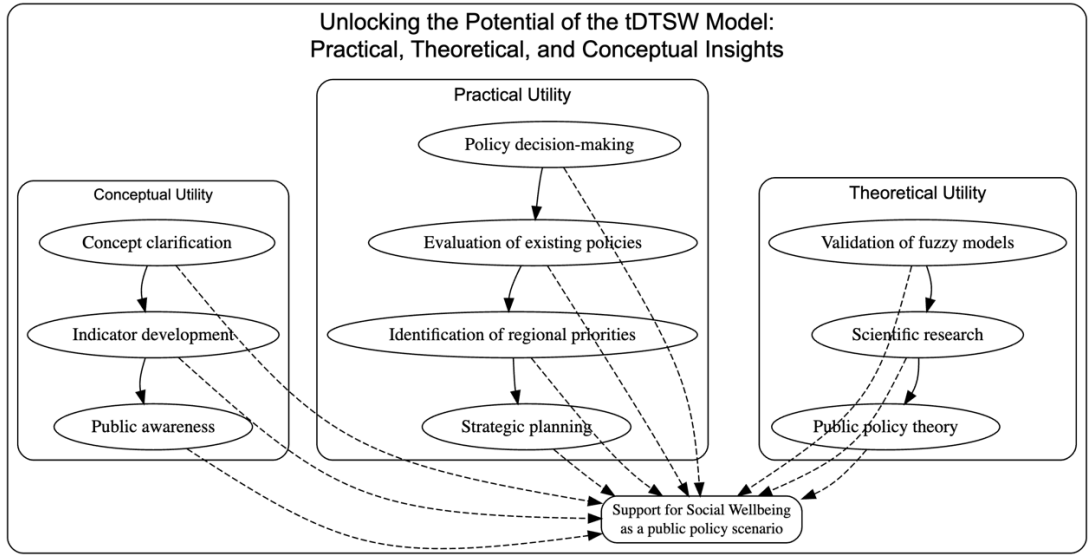

Source: Own elaboration

**4.4.2 Theoretical utility**

*Validation of fuzzy models:* This analysis validates the applicability of fuzzy models in decision-making. It demonstrates how membership concepts and rules can be applied in real-world situations.

*Scientific research:* The model (tDTSW) and its analysis provide a theoretical foundation for further research in fields such as fuzzy systems theory, political science and the economics of well-being.

*Public policy theory:* It helps develop and refine theories about how democracy and transparency influence social well-being, which can have a significant impact on public policy formulation.

**4.4.3 Conceptual utility**

*Concept clarification:* The model (tDTSW) clarifies the concepts of democracy, transparency and social well-being by showing how they are related in specific situations.

*Indicator development:* The analysis can inspire the development of concrete indicators that measure the quality of democracy, transparency and social well-being, which is essential for policy monitoring and evaluation.

*Public awareness:* By explaining how these factors relate to each other, public awareness of the importance of democracy and transparency in improving social well-being can be increased.

## 4.5 Support for Social Wellbeing as a public policy scenario

This analysis lends strong support to the adoption of public policies aimed at improving the overall welfare of communities. By gaining insights into how democracy and transparency impact social well-being across diverse scenarios, policymakers can craft more potent strategies for advancing societal welfare.[65] Additionally, by furnishing a robust theoretical groundwork and clarifying these core concepts, this analysis aids in shaping more enlightened policies and encourages increased public involvement in decisions that shape the well-being of society. Ultimately, this has the potential to drive lasting enhancements in the quality of life for both individuals and communities.

## 4.6 Preliminary conclusions

In Chapter 4 of our tDTSW model delves deep into the fuzzy inference process, illuminating the intricate relationships between Democracy (D), Transparency (T) and Social Wellbeing (SW). Through nine distinct scenarios, we unveil the varying degrees of influence these variables exert on each other. The graphical representation with color-coded lines provides a precise mathematical depiction of these relationships, aiding in rigorous interpretation.

Practically, this analysis can empower policymakers to make informed decisions, evaluate existing policies and identify regional priorities for improving social well-being. Theoretically, it validates the applicability of fuzzy models and contributes to research in various fields. Conceptually, it clarifies core concepts and promotes public awareness.

The research highlights the need of establishing policies targeted at enhancing the social well-being, which presents a chance to improve the lives of both people and whole communities. This chapter acts as a foundational piece, laying the groundwork for the actual implementations and real-world.

---

[65] See. Pena-Trapero, B. (2021). La medición del Bienestar Social: una revisión crítica. Studies of Applied Economics. Vol. 27 No. 2 (2009). https://ojs.ual.es/ojs/index.php/eea/article/view/4919

# Conclusion

Our deep dive into the intricate interplay between democracy, transparency and social welfare has been a profound voyage marked by invaluable insights and eye-opening discoveries. As we draw the curtains on our extensive examination, it is undeniably evident that these principles transcend mere theoretical notions; they are the fundamental core of a righteous and enduring society.

In the inaugural chapter of our expedition, we set the stage by underscoring the unparalleled importance of transparency and democracy as the dual foundations upon which to erect societies that are fairer, more just, and sustainable.

Our model, tDTSW, offers a framework to understand and navigate the complexities of these principles, providing valuable insights for decision-makers and researchers alike. Our examination of Finland, Singapore and New Zealand's governance models served as inspiring case studies. These nations have showcased the transformative power of transparent and accountable governance. Their success stories illustrate the tangible benefits of upholding democratic principles and transparency consistently. We also confronted the formidable challenges that modern democracies face, from the tension between capitalism and ecological sustainability to the underrepresentation of women in leadership roles.
 However, within these challenges, we discovered opportunities. Education emerges as a potent tool for nurturing informed and engaged citizens, while support for women's organizations becomes a catalyst for achieving gender equality in leadership.

In Chapter 2, the tDTSW framework provided a comprehensive perspective on the intricate dynamics between democracy, transparency, and social wellbeing within the context of politics and society. This framework underscores the central role of democracy as a catalyst for positive change and highlights the critical significance of transparency, accountability, trust, and effective governance in enhancing the overall welfare of citizens. Beyond its theoretical value, the tDTSW framework extends its practical implications to guide policymakers and stakeholders. It emphasizes the imperative of transparent governance, the necessity of reform implementation and the importance of nurturing trust to enhance the efficacy of democratic systems and elevate societal welfare.

Chapter 3 introduced fuzzy logic as a powerful framework for assessing the complex relationship between democracy, transparency, and social wellbeing. This innovative approach allowed us to handle imprecise and uncertain information, acknowledging the inherent vagueness in real-world phenomena. The tDTSW model, with its fuzzy sets, membership functions and fuzzy rules, provides a structured method for understanding how democracy and transparency interact to influence social wellbeing across various scenarios.

In Chapter 4, we delved even deeper into the fuzzy inference process, unveiling the varying degrees of influence these variables exert on each other. The graphical representation with color-coded lines offered a precise mathematical depiction of these relationships, aiding in rigorous interpretation. This analysis equips policymakers with the tools to make informed decisions, evaluate existing policies and identify regional priorities for improving social wellbeing.

As we conclude our journey, it is evident that democracy and transparency are not just ideals; they are actionable principles that can shape the destiny of societies. Our model, tDTSW, offers a framework to understand and navigate the complexities of these principles, providing valuable insights for decision-makers and researchers alike. The journey towards more equitable, sustainable, and just societies demands unwavering dedication, adaptability and a collective effort involving governments, civil society, the private sector, and every individual. We must persist in upholding these principles as our moral compass, guiding us toward a brighter and more inclusive future for all.